\documentclass[aps,prb,twocolumn,floatfix,superscriptaddress,10pt]{revtex4-1}

\usepackage{graphicx}
\usepackage{amsmath}
\usepackage{amsfonts}
\usepackage{amssymb}

\begin{document}

\bibliographystyle{unsrt}

\title{Collective excitations and low temperature transport properties
of bismuth}
\author{P. Chudzinski}
\affiliation{DPMC-MaNEP, University of Geneva, 24 Quai
Ernest-Ansermet CH-1211 Geneva, Switzerland}
\author{T. Giamarchi}
\affiliation{DPMC-MaNEP, University of Geneva, 24 Quai
Ernest-Ansermet CH-1211 Geneva, Switzerland}

\begin{abstract}
We examine the influence of collective excitations on the
transport properties (resistivity, magneto-optical conductivity)
for semimetals, focussing on the case of bismuth. We show, using
an RPA approximation, that the properties of the system are
drastically affected by the presence of an acoustic plasmon mode,
consequence of the presence of two types of carriers (electrons
and holes) in this system. We find a crossover temperature $T^*$
separating two different regimes of transport. At high
temperatures $T > T^*$ we show that Baber scattering explains
quantitatively the DC resistivity experiments, while at low
temperatures $T < T^*$ interactions of the carriers with this
collective mode lead to a $T^5$ behavior of the resistivity. We
examine other consequences of the presence of this mode, and in
particular predict a two plasmon edge feature in the
magneto-optical conductivity. We compare our results with the
experimental findings on bismuth. We discuss the limitations and
extensions of our results beyond the RPA approximation, and
examine the case of other semimetals such as graphite or
$1T-TiSe_{2}$.
\end{abstract}

\maketitle

\section{Introduction}

Bismuth is a material which plays an important role in solid state
physics. Due to an extremely small Fermi surface this material
provides the remarkable possibility to observe strong effects
induced by the presence of external fields, pressure and
temperature, even if these external forces are of moderate
amplitude. Several important phenomena, Shubnikov-de
Haas\cite{SdH_classic30} and de Haas-van
Alphen\cite{dHvA_classic30} effects, were observed for the first
time in bismuth. In a last few years a series of experiments have
once again drawn the attention of the community to elemental
bismuth, and challenged our understanding of this material. High
pressure optical spectroscopy measurements indicate that the
mechanism of semimetal to semiconductor decay is not fully
understood\cite{Tediosi_pressure}. Reflectivity
measurements\cite{Tediosi_plasmaron07} showed large changes in the
plasmon frequency and anomalous mid-infrared features, indicating
strong scattering of the electronic degrees of freedom by a
plasmon collective mode.  An extremely strong Nernst signal with
unusual temperature dependence (both at
low\cite{Uher_Nernst_e-h78} and high fields\cite{Behnia_Nernst07,
Behnia_BiSb}) was also reported.

These recent experiments were showing clearly that the question of
transport in bismuth was still not understood. In fact similar
questions still existed for the standard resistivity as well. The
majority of the resistivity experiments had been done in late 70s,
beginning with the works of
Hartmann\cite{Hartmann_experim-full69},
Kukkonen\cite{Kukkonen_experim77} and later
studies\cite{Kraak_highT82}, who showed that down to 4K the Fermi
liquid theory (with components of very different masses) works
quite well. The $T^2$ resistivity behavior at lowest temperatures
was explained within this
theory\cite{Kukkonen_Baber76,Kukkonen_experim77}. However the
discussion was not closed because one year later more detailed,
lower temperature data by Uher\cite{Uher_Bi-rho77} was published,
and showed a significant deviation from expectations: the $T^2$
behavior changes smoothly into a $T^5$ behavior at the lowest
temperatures. A quite complex theory involving coupling to
particular group of phonons was proposed as an explanation of this
result\cite{Kukkonen_el-phon78}.

The purpose of this work is to re-examine the theory of transport
in semimetals. We show that these anomalous transport properties
have a simple explanation. They come from the fact that the
coupling between electrons and holes with very different masses
induces many body corrections to the Fermi liquid picture. In fact
the above mentioned change in the resistivity was the first
example how interactions in semimetals modify the simple Fermi
liquid picture. Although our study is mostly focused on bismuth,
we also examine other materials which have been recently the
subject of intensive studies, such as
graphite\cite{Maude_graphite09, Behnia_graphite09} and
$1T-TiSe_2$\cite{Aebi_EI07, Tutis_SC09}.

The structure of this paper is as follows. In Sec.~\ref{sec:mecha}
we introduce the model of interaction between the electrons and holes.
We show that at low temperatures, a collective acoustic plasmon
mode exists and plays a central role in the properties of the material.
We discuss also the high temperature regime of conductivity and show
that the Baber mechanism\cite{Baber_classics37} is the dominant source
of resistivity in this regime. We then examine in Sec.~\ref{sec:lowt}
the low temperature regime for the resistivity. We develop an effective theory
for this regime and derive a new $\rho(T)$ dependence, which is a direct
consequence of the existence of a collective acoustic plasmon mode.
We examine the magneto-transport in Sec.~\ref{sec:magnetop}. We show
in particular that a double plasma edge must exist. Some discussion
on the validity of the approximations used to derive the above mentioned
results are indicated in Sec.~\ref{sec:discuss} and conclusions in Sec.~\ref{sec:conclusion}.
Some technical details can be found in the appendices.

\section{Mechanism of resistivity} \label{sec:mecha}

\subsection{Band structure and hamiltonian}

\begin{figure}
  \vspace{3cm}
  \centerline{\includegraphics[width=0.9\columnwidth]{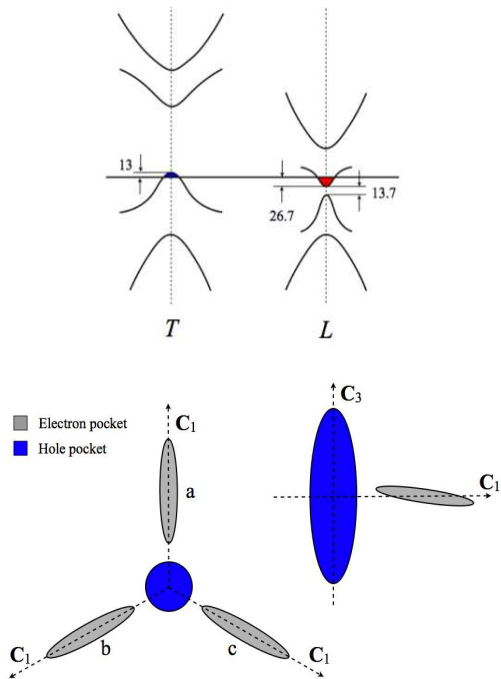}}
  \caption{\label{fig:band-structure} A sketch of the band structure of bismuth: three electron pockets and
  a central hole pocket. (top) the energy dispersion. Energies are in meV. (bottom)
  The four pockets. The scale is grossly exaggerated. In fact
  $k_{F}$ are $10^4$ smaller than intra-pocket distances. The electron pockets are
  slightly tilted out of bisectrix-binary plane $C_1-C_2$ perpendicular to the trigonal axis $C_3$. Figure after Ref.\onlinecite{Ted_thesis}}
\end{figure}

The peculiarity of bismuth comes from the very small
characteristic energy scales of its Fermi liquid (see
Fig.~\ref{fig:band-structure}).
This stems from a slightly distorted
cubic crystal structure (the distortion angle is smaller
than $3^\circ$). In the absence of distortion bismuth would
be a band insulator, instead it is a quite rare
rhombohedral space group $A7$ (without inversion symmetry).
Bismuth becomes a semimetal with very small amounts of fermions
active at accessible energies.

The hamiltonian of those carriers close to Fermi energy in bismuth
reads in general:
\begin{equation}\label{eq:ham_gen}
    H=H_{0}^{h}+\sum_{\nu=1}^3 H_{0}^{e\nu}+H_{int}^{h-h}+H_{int}^{e-e}+H_{int}^{e-h}
\end{equation}
In the above $h$ denotes a hole pocket, while there are three electron pockets denoted
by $e\nu$. The first two terms are the free fermion kinetic energies. We approximate
the kinetic energy of each type of carrier by a a free dispersion relation
\begin{equation}
 \xi^\alpha(k) = \sum_{i=1}^3 \frac{\hbar^2 k_i^2}{2 m_i^\alpha} - E_F^\alpha
\end{equation}
where $\alpha$ denotes the species, the index $i$ are the three principal axes of the energy
ellipsoid (see Fig.~\ref{fig:band-structure}), $m_i^\alpha$ is the mass tensor, the $k$
are the momenta centered on the corresponding pocket and $E_F^\alpha$ the Fermi energy of the corresponding
species. The values of the parameters are summarized in
Table~\ref{tab:masses}.
\begin{table}
\begin{center}
 \begin{tabular}{||c||c|c|c|c|c|c||}
    \hline
     & $m_{1}$ & $k_{F1}$ & $m_{2}$ & $k_{F2}$ & $m_{3}$ & $k_{F3}$ \\
    \hline
    holes & 0.067 & 0.01 & 0.067 & 0.01 & 0.612 & 0.03  \\
    electrons & 0.198 & 0.06 & 0.0015 & 0.005 & 0.0021 & 0.005  \\
    \hline
 \end{tabular}
 \end{center}
 \caption{\label{tab:masses} The band parameters of bismuth according to Ref.~\onlinecite{Allen_Bi-band95}.
 Masses are in units of the mass of the electron, and wavevectors in units of reciprocal length $1.386 \AA ^{-1}$.
 The values are given along the principal axes of the ellipsoids as defined in Fig.~\ref{fig:band-structure}.
 }
\end{table}
We want to emphasize that this peculiar band structure was computed
quite accurately and this result was later confirmed by many
experimental probes.

The latter terms in (\ref{eq:ham_gen}) are interactions, whose influence is the subject
of this study. We divided interactions into three groups:
hole-hole, electron-electron and electron-hole. Each one of these terms has the form
\begin{equation} \label{eq:inter_gen}
 H_{int}^{\alpha-\beta} = \sum_{q} V^{\alpha\beta}(q) \rho_\alpha(q) \rho_\beta(-q)
\end{equation}
where $\alpha,\beta$ runs among the species. The density operators $\rho_\alpha(q)$ is given by
\begin{equation} \label{eq:densdef}
 \rho_\alpha(q) = \sum_k c^\dagger_{\alpha,k+q} c_{\alpha,k}
\end{equation}
where the $c^\dagger,c$ are the standard fermionic creation and destruction operators and a summation overs the spin degrees of freedom
is implicit. The interaction potential is the long-range Coulomb potential
\begin{equation} \label{eq:coulomb}
 V^{\alpha\beta}_{\rm coul}(q) = \frac{e_\alpha e_\beta}{\epsilon_{\infty} q^2}
\end{equation}
where the hole charge is the opposite of the electronic one $e_h =
-e_e$ and $\epsilon_{\infty}$ is the dielectric constant due to
the rest of the material. Because of the very small size of the
pockets and the large distance in momentum space among them all
interactions which implies a transfer of particle from one pocket
to the other must occur with a large momentum transfer. These
terms are thus potentially strongly suppressed compared to the
intra-pocket interactions given the long-range nature of the
Coulomb potential (\ref{eq:coulomb}).

\subsection{Screened Coulomb interaction: acoustic plasmon} \label{sec:scoul}

The smallness of the bismuth Fermi surface shown in
Fig.\ref{fig:band-structure} and incorporated in the first two
terms of the hamiltonian (\ref{eq:ham_gen}) has certain
non-trivial implications for the transport properties.
When $k_{F}^{-1}\approx 10^{4}\AA$ the interaction with
angstrom size impurities is strongly reduced. The probability
of intra-pocket umklapp scattering is also very low. Since the Debye
temperature for Bi is around $150K$ it implies that at $T\sim
1K$ phonon influence will be rather weak. Moreover thermal
transport is proven experimentally to be ballistic. No evidence of
any additional order parameter, generating quasi-particles on
which carriers could scatter was found in the many decades of
careful studies of this element. Thus in order to understand the transport
properties we can restrict ourselves to
the Fermi liquid part of the problem described by the hamiltonian
(\ref{eq:ham_gen}).

Given the above points, the main type of scattering entering resistivity
should be the so-called Baber scattering\cite{Baber_classics37}, coming from the presence of several
types of carrier of different masses. In contrast with all the above mentioned
scattering processes Baber scattering is strongly favored by the
band structure of Bi -- the ratio of electrons and holes masses
being in some directions more than a factor of $10$. The resistivity in Bismuth
is thus linked directly to the electron-hole interaction term in (\ref{eq:ham_gen}).

\begin{figure}
  \includegraphics[width=0.99\columnwidth]{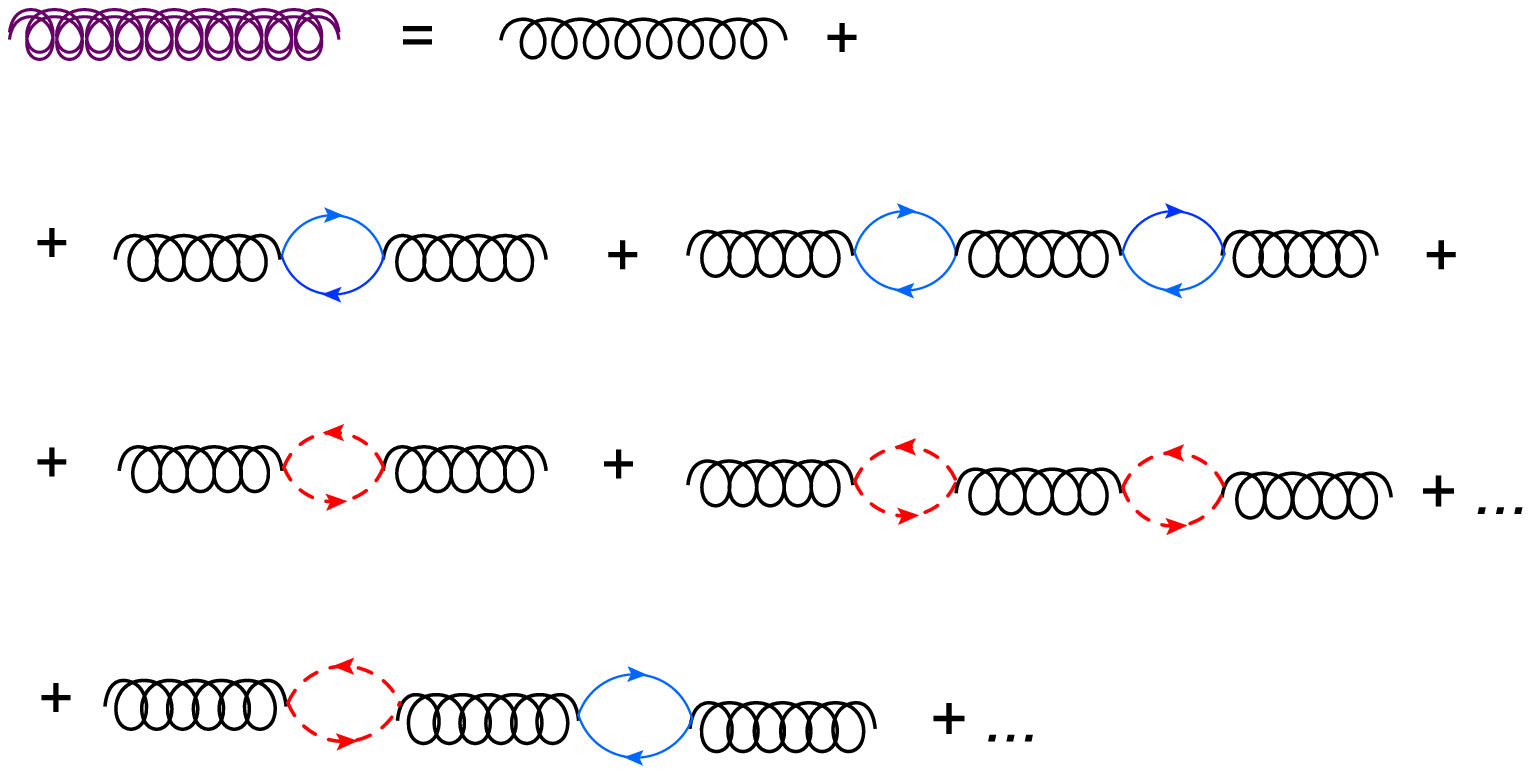}\\
  \caption{\label{fig:rpa_diag} The RPA series for effective interactions for the case of a two component system.
  We denote generically these components by $r$ (rapid or fast) and $s$ (slow). In the
  above diagram, the thick coil is the screened effective interaction $V_{\rm eff}(q,\omega)$, the solid blue denotes using the standard diagrammatic
  convention a fast particle, while the doted red line is a slow particle. The black coil is the bare Coulomb potential (\ref{eq:coulomb}).}
\end{figure}

Although the bare form of such a term is given by (\ref{eq:inter_gen}), the presence of the
two species of carriers leads to a strong renormalization of the bare Coulomb interaction due to
dynamical screening. We thus need to examine the effects of screening on the interaction potential
$V^{\alpha\beta}(q)$. Normally we have a tensor structure for the screened interaction
(or alternatively the dielectric constant), but if we assume that the longitudinal and transverse
modes do not mix, which is a good approximation in Bismuth, then we can use the standard
Random Phase Approximation (RPA) for the Coulomb potential. In the presence of two types of carrier
the RPA equations are indicated in Fig.~\ref{fig:rpa_diag}.

The equation of Fig.~\ref{fig:rpa_diag} can be solved doing first the re-summation on one of the species:
\begin{equation} \label{eq:screened_coul}
\tilde{V}^{ee}(q,\omega) = \frac{V^{ee}_{\rm coul}(q)}{1 - V^{ee}_{\rm coul}(q) \Pi_{hh}(q,\omega)}
\end{equation}
where $\Pi_{hh}(q,\omega)$ is the retarded density-density correlation function for the holes, including
the interactions which are not already present in the RPA chain of bubbles of Fig.~\ref{fig:rpa_diag}.
The simplest and usual approximation for $\Pi_{hh}$ consists in taking the free correlation $\Pi^0_{hh}(q,\omega)$. Using the Hamiltonian
(\ref{eq:ham_gen}) one has
\begin{equation} \label{eq:lindsum}
 \Pi^0_{hh}(q,\omega) = \frac1\Omega \sum_k \frac{f(\xi(k)) - f(\xi(k+q))}{\omega + \xi(k) - \xi(k+q) + i \delta}
\end{equation}
where $\Omega$ is the volume of the system, $f$ the Fermi factors, $\delta = 0^+$.
Using this expression one can get the screening of the electron-electron potential
\begin{equation} \label{eq:effint}
 V_{\rm eff}^{ee}(q,\omega) = \frac{\tilde{V}^{ee}(q)}{1 - \tilde{V}^{ee}(q,\omega) \Pi_{ee}(q,\omega)}
\end{equation}
where $\Pi_{ee}(q,\omega)$ denotes the sum over the three electron
pockets of the corresponding retarded density-density correlation
$\Pi_{ee}(q,\omega) = \sum_{\nu=1}^3 \Pi_{ee,\nu}(q,\omega)$. The
expression (\ref{eq:effint}) can be rewritten as
\begin{equation} \label{eq:collecmode}
  V_{\rm eff}^{ee}(q,\omega) = \frac{V^{ee}_{\rm coul}(q)}{1 - V^{ee}_{\rm coul}(q)[\Pi_{hh}(q,\omega) + \Pi_{ee}(q,\omega)]}
\end{equation}
where in a similar way than for the holes one can approximate the correlation function by its free (or Fermi liquid) value.

As can be seen from (\ref{eq:collecmode}) a strong resonance in $V_{\rm eff}^{ee}(q,\omega)$ exists when the denominator is small,
which indicated the presence of a collective mode. As usual the dispersion relation of the mode is given by
\begin{equation}\label{eq:eps=0cond}
 1 = V^{ee}_{\rm coul}(q){\rm Re}[\Pi_{hh}(q,\omega) + \Pi_{ee}(q,\omega)]
\end{equation}
where ${\rm Re}$ denotes the real part. The imaginary part gives
the damping of the mode. As is well
known\cite{Pines_ac-plasm-classic62, Ruvalds_ac-plasm-review81,
Cottey_ac-plasm-basic85} for the case of a two component system
(\ref{eq:eps=0cond}) has two solutions. One is the standard
optical plasmon, but a second one is an acoustic mode
$$\omega_{ac}=uq=(c_{\rm ac}- \imath \tau^{-1})q$$
where $Re[u]=u'=c_{\rm ac}$ is the acoustic plasmon velocity and
$Im[u]=u''=-\tau^{-1}$ the damping of the mode. For the case of
isotropic Fermi surfaces with the same Fermi wavevector for the
two species (neutrality condition) and one heavy $m_s$ and light
masses $m_r$ (for slow and rapid carriers respectively) such that
$(V_{Fs}/V_{Fr})^2 \ll 1 \ll m_{s}/m_{r}$, where the $V_{Fi}$
($i=r,s$) are the respective Fermi velocities $k_{Fi}/m_i$ with
$E_F^i = k_{Fi}^2/(2m_i)$ in the zeroth order approximation
(extreme mass difference), one has
\begin{equation}\label{eq:velocity}
 c_{\rm ac}^{0}=\sqrt{V_{Fr}V_{Fs}/3}
\end{equation}
and
\begin{equation}\label{eq:damp}
    (\tau^{0})^{-1}= \pi V_{Fs}/12
\end{equation}
Physically such an acoustic plasmon comes from the fact that for
two different velocities, one of the species (the light one) will
have a much larger Fermi velocity than the acoustic plasmon mode.
In that case the corresponding $\Pi_{rr}(q,\omega)$ essentially
tends to a constant $\Pi_{rr}(q\to0,\omega=0)$ which is the
density of states at the Fermi level. This mode is fast enough to
screen the Coulomb interaction. (\ref{eq:screened_coul}) would
just become
\begin{equation} \label{eq:inter1}
\tilde{V}^{rr}(\omega = c_{\rm ac} q \to 0) = \frac{1}{-
\Pi_{rr}(\omega = c_{\rm ac} q \to 0)}
\end{equation}
transforming the original long range interaction into a short range one. Using the well known result for the Lindhard function one gets
\begin{equation}
 - \Pi_{rr}(\omega = c_{\rm ac} q\to 0) = \frac{m_r k_{Fr}}{\pi^2} + \imath \frac{m_r^2 c_{\rm ac}}{2\pi}
\end{equation}
For the heavy mode one is in the opposite limit $c_{\rm ac} \gg
V_{Fs}$ for which the Lindhard function is approximately
\begin{equation}
  - \Pi_{ss}(\omega = c_{\rm ac}q \to 0) = \frac{m_s k_{Fs}}{\pi^2} \frac{k_{Fs}^2}{3 c_{\rm ac}^2 m_s^2}
\end{equation}
Substituting (\ref{eq:inter1}) into (\ref{eq:effint}) one gets the results (\ref{eq:velocity}) and (\ref{eq:damp}).

Such collective excitation has been already investigated in
several systems \cite{Pines_ac-plasm-classic62,
Ruvalds_ac-plasm-review81,Bennacer_ac-plasm-Bi89} but is normally
very elusive. It exists usually very close to the single-particle
excitation spectrum, so it can be easily Landau
\emph{overdamped}. There were proposals\cite{Das_Sarma_ac-plasm99} suggesting its
measurability in artificial 1D systems (quantum wires), but even
then its intensity was shown to be much weaker than optical
plasmon and was easily suppressed by disorder. However we claim in
the present paper that such a collective mode plays an important
role in Bismuth. Computing precisely its parameter is a singularly
complicated calculation given the complexity of the Fermi surface
and the presence of the various masses. Nevertheless we give in
Appendix~\ref{ap:bismuth_plasmon} arguments for the existence of
such a mode. Mostly we will proceed along the lines that such a
mode exists and explore the consequences for the various transport
properties.

Note that the mode never exists as a perfectly sharp mode, because
of the damping (\ref{eq:damp}) which always gives a finite width
of the order of $\tau^{-1}$ to this acoustic plasmon resonance. In
addition the mode will completely disappear when its dispersion
relation will enter the the particle-hole continuum of excitations
of the heavy (slow) species. Beyond this point the mode is
severely damped and does not exist as a collective excitation
anymore. The corresponding wavevector $q^{*}$ can be determined by
matching the energy of the acoustic plasmon $\omega = c_{\rm ac}
q$, taking into account a width of order $\tau^{-1} q$ with the
edge of the particle-hole spectrum $q(q+2k_{Fh})/2m_{h}$. An
estimate of this wavevector is thus
\begin{equation} \label{eq:upper}
 q^{*} \simeq 2 m_h [c_{\rm ac} - V_{Fh} - \tau^{-1}]
\end{equation}
This condition which is essentially similar to the one for $k_c$
in Ref.~\onlinecite{Bennacer_ac-plasm-Bi89} (where $V_{Fs}q$ and $2k_{Fs}$
were taken as energy and momentum units).

Correspondingly $q^{*}$ defines an energy and temperature scale
\begin{equation} \label{eq:debplas}
 T^{*} \sim c_{\rm ac} q^{*}
\end{equation}
This temperature play a similar
role than the Debye temperature for acoustic phonons: there are no
bosonic states to be occupied beyond. On the other hand below this
temperature the physics of the system is affected by the existence
of the extra collective mode -- there are plasmon states whose
occupation fluctuations can affect carriers mobilities. Based on
the above we see that we can potentially distinguish two very
different temperature regimes:

\paragraph{$T>T^{*}$}
This is the high temperature (HT) regime. In that regime there is
no collective mode and we can treat Bismuth as a double Fermi
liquid. We can thus deal with the electron-hole scattering in the
usual way, for example by solving the Boltzman transport equation,
or equivalent approximations. We will briefly discuss this rather
conventional regime in Sec.~\ref{ssec:standard_Baber}.

\paragraph{$T<T^{*}$}
This is the low temperature (LT) regime. In that regime, the
presence of the acoustic plasmon plays a central role in the
interactions among particle. It will thus affect strongly the
scattering between electrons and holes and lead to very different
transport properties. The study of the consequences of such a mode
is at the heart of the present paper and will be done for the
resistivity in Sec.~\ref{sec:lowt} and in the subsequent
sections for other transport coefficients.

\subsection{High temperature regime: Baber resistivity}\label{ssec:standard_Baber}

Let us now examine the resistivity itself. As discussed above we have to distinguish two different
regimes of temperature. We examine in this section the high temperature one, which shows relatively
conventional transport properties, and will concentrate on the low temperature regime in the next section.

At high temperature no collective mode is present and thus we have two well defined
Fermi liquids. The resistivity is coming from the electron-hole interaction since this
is the only term that does not conserve the total current. Indeed given the quadratic dispersion
relation the current is proportional to the momentum and thus the intra-species interactions
conserve the total current. The main source of resistivity is this regime is the Baber scattering\cite{Kukkonen_Baber76},
which leads within a standard Boltzmann approximation to
\begin{equation}\label{eq:rho_Baber_standard}
    \rho=(\frac{m_{e}m_{h}k_{B}}{3\pi e\hbar^3 \sqrt{n}})^2 W T^2 = A T^2
\end{equation}
where $W$ is the e-h scattering rate and $n$ the density of particles of a given species. This gives an approximation
for the coefficient $A$ of the $T^2$ term in the resistivity $\rho=AT^2$, which is usually
the quantity extracted from experimental data. Thus, provided that
one is able to compute $W$, we have a parameter free fit for
experiment. The prefactor $W$ was computed using a Thomas-Fermi approximation
for the Coulomb interaction\cite{Kukkonen_Baber76} leading to
\begin{equation}\label{eq:scatt_amplit}
    W=\frac{2\pi}{\hbar}\int_{0}^{2k_{Fs}}\frac{dq}{2k_{Fs}}V_{\rm TF}(q)^2[3(q/2k_{Fs})^2]
\end{equation}
where $k_{Fs}$ is the Fermi wavevector of the slower component.
$V_{\rm TF}(q)$ is the Thomas-Fermi screened Coulomb
interaction:
\begin{equation}\label{eq:potential_Kukk}
    V_{eff}(q)=\frac{4\pi e^2}{\epsilon_{\infty}q^2 +\kappa_{T-F}^{e2}+\kappa_{T-F}^{h2}}
\end{equation}
Ref.~\onlinecite{Kukkonen_Baber76} estimated the above terms using
$\epsilon_\infty=100$,
$\kappa_{T-F}^{i}=3(4m_{i}^2 k_{Fi}/\pi\hbar^2)$
where $m_e=0.03$, $m_h=0.15$ which were known from previous
independent measurements. This led to $A = 8n\Omega cm K^{-2}$ in
good agreement with the experimental data of that time $8n\Omega
cm K^{-2}$ (the first reported in
Ref.\onlinecite{Fenton_rhoBi67}), $14.5n\Omega cm K^{-2}$ (in
Ref.\onlinecite{Hartmann_experim-full69}) and also the more recent
ones\cite{Uher_Bi-rho77, Behnia_Nernst07} $12n\Omega cm K^{-2}$
for measurements along the binary axis. With the same formulas an
even better value $A=14n\Omega cm K^{-2}$ can be found if we use
more recent values of parameters: $\epsilon_{\infty}=88$, $m_e =
0.04$, $m_h = 0.14$. The $Bi_{1-x}Sb_{x}$ compound, with $x=0.037$
($n_{x=0.037}\approx n_{x=0}/3$) was also
measured\cite{Behnia_BiSb} and the experimental $A=33n\Omega cm
K^{-2}$, is in agreement with the expectations from
(\ref{eq:rho_Baber_standard}), showing that the Baber scattering
is indeed the good description of the the transport in this regime
of parameters. However low temperature deviation from this
simplistic picture were reported already in early works (see
Ref.\onlinecite{Kukkonen_experim77} and a detail discussion there)

Within the framework of the Baber scattering, one can go beyond the traditional Baber formula
(\ref{eq:rho_Baber_standard}) in two ways. First the large temperature behavior can be computed as well\cite{Giamarchi_Shastry92}.
At high enough temperature the $T^2$ behavior of (\ref{eq:rho_Baber_standard}) crosses over,
at a temperature of the order of $T_{l} \sim 0.2 \sqrt{T_{Fe}T_{Fh}}$ to a linear temperature
dependence. Experimental data in Bismuth shows indeed above $T = 3K$ a clear $T^2$
dependence, becoming linear above $T=25K$. This is in reasonable agreement with the
estimate given by the above formula which for Bismuth would be $T_1 \sim 35K$.
Second, one can refine the calculation of the coefficient $A$ by taking into account
the ellipsoidal character of the Fermi surface, rather than using the best fit for spherical Fermi
surface approximation ($m_{e}=0.04$, $m_{h}=0.14$). The procedure is given in
Appendix~\ref{ap:ratioapp} for the simplified case of two pockets only.
The ratio of resistivities in different directions $x$ and $z$ is proportional
to:
\begin{equation}\label{eq:ratio}
    \frac{\rho_{x}}{\rho_{z}}=\left[\frac{|m_{ex}-m_{hx}|/(m_{ex}+m_{hx})}{|m_{ez}-m_{hz}|/(m_{ez}+m_{hz})}\right]^2
\end{equation}
Given the masses of Table~\ref{tab:masses} we expect that resistivity will be largest along the trigonal axis
of the crystal. The value of the asymmetry $\approx 1.4$ that the above formula gives
is slightly larger than the experimentally\cite{Hartmann_experim-full69} found one $\approx 1.15$.
However one has to remember that due
to highly anisotropic ellipsoids constituting the Fermi surface
the experiment requires a very precise monocrystal orientation.

However the Baber scattering does not allow to understand the low temperature regime which in Bismuth
corresponds to $T < 3K$. For this regime one has to invoke the existence of the acoustic plasmon. This is the
regime that we consider now.

\section{Low temperature regime} \label{sec:lowt}

\subsection{Effective hamiltonian for lowest temperatures}\label{sec:acous_res}

As we determined in the previous section, at low enough temperature
an acoustic plasmon mode exists. The very existence of this mode
implies that for some values of the frequency and momenta the effective
interaction $V_{\rm eff}(q,\omega)$ between the carriers will have a pole.
One can thus expect this pole to dominate the low temperature transport.

\begin{figure}
 \vspace{3cm}
  \begin{center}
   \includegraphics[width=0.9\columnwidth]{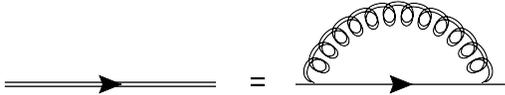}\\
  \end{center}
  \caption{\label{fig:electr_self-energy} a) The self-energy of a fermion (an electron or hole) experiencing $V_{eff}(q,\omega)$ interaction
  (a double coil as in Fig.~\ref{fig:rpa_diag}) during its propagation; as explained in the text the effective interaction $V_{eff}(q,\omega)$ is of the resonance form.}
\end{figure}

In order to analyze the consequences of such a pole, we derive an effective
low energy Hamiltonian for which we consider this collective mode as a
particle. This can be done by replacing the term corresponding to
$V_{\rm eff}(q,\omega)$ (as shown in Fig.~\ref{fig:electr_self-energy})
by the propagator of a particle, representing the bosonic fluctuations of the plasma and ensuring that terms such as the self-energy
terms such as the ones of Fig.~\ref{fig:electr_self-energy} are correctly reproduced.
Such a model is given by the coupling of electrons and holes to phonon-like excitations
\begin{equation} \label{eq_hphonon}
 H = H_0^h + H_0^e + \sum_q \omega_q b^\dagger_q b_q +
  \frac1{\sqrt{\Omega}} \sum_{\alpha=e,h}\sum_q M_q^\alpha [b^\dagger_{-q} + b_q] \rho_\alpha(q)
\end{equation}
with such an Hamiltonian the electron self energy of Fig.~\ref{fig:electr_self-energy} would be
\begin{equation} \label{eq:selfphon}
 \Sigma_e(q,i\nu_n) =  -\frac1{\beta\Omega} \sum_{\omega_n,q} (M_q^{e})^2 D(q,i\omega_n) \frac1{i\nu_n + i\omega_n - \xi_e(k+q)}
\end{equation}
where the $\nu_n$ and $\omega_n$ are the usual Matsubara frequencies and $D(q,i\omega_n)$ is the phonon propagator
\begin{equation}
D(q,i\omega_n) = - \frac{2\omega_q}{\omega_n^2 + \omega_q^2}
\end{equation}
The corresponding term for the original Hamiltonian is shown in Fig.~\ref{fig:electr_self-energy} and is given by
\begin{equation}
 \Sigma_e(q,i\nu_n) =  -\frac1{\beta\Omega} \sum_{\omega_n,q} V_{\rm eff}(q,i\omega_n) \frac1{i\nu_n + i\omega_n - \xi_e(k+q)}
\end{equation}
One can show on this term, and the other diagrams, that the expansion is essentially identical provided one identify the proper ``phonon'' propagator
and interaction vertex $M_q$. This can easily be done on the spectral function $A(q,\omega) = -\frac1\pi {\rm Im}D(q,i\omega_n \to \omega + i \delta)$. We only consider $\omega > 0$ since $\omega < 0$ can easily be deduced from it. The ``phonon'' spectral function is
\begin{equation}
 A_{\rm ph}(q,\omega >0) = \delta(\omega-\omega_q)
\end{equation}
The effective potential $V_{\rm eff}(q,\omega)$ of (\ref{eq:collecmode}) is more difficult to evaluate fully. However using the simple case of two types of particles
(fast and slow) as discussed in the Sec.~\ref{sec:scoul} one has
\begin{equation}
 V_{eff}(q,\omega) = \left(\frac{3\pi^2 m_r}{k_{Fr}^3}\right) \frac1{c_{\rm ac}^2}{1- c_{\rm ac}^2 q^2/\omega^2 + \imath \pi c_{\rm ac}/2}
\end{equation}
There is a finite lifetime due to the imaginary part (\ref{eq:damp}), leading to a lorentzian spectral function. If for the moment we ignore this finite width and assimilate
the resonance to a $\delta$-function peak we have for the spectral function
\begin{equation}
 A_{\rm plasmon}(q,\omega >0) = \left(\frac{3\pi^2 m_r}{k_{Fr}^3}\right) \frac{c_{\rm ac}^3 |q|}2 \delta(\omega - c_{\rm ac} |q|)
\end{equation}
One can thus directly identify the two processes with
\begin{equation} \label{eq:ident}
\begin{split}
 \omega_q &= c_{\rm ac} |q| \\
 M_q^{e0} &=  \left(\frac{3\pi^2 m_r c_{\rm ac}^3}{2 k_{Fr}^3}\right)^{1/2} |q|^{1/2}
\end{split}
\end{equation}
where $M_q^{e0}$ is the zeroth order approximation for the
strength of electron-plasmon interaction (as introduced in
Eq.~\ref{eq_hphonon}). The above identification allows us to
replace the problem of the screened Coulomb potential, by a
problem of electrons and holes interacting with a bosonic
particle. Of course this particle represents the quantization of
the acoustic plasmon.

Let us make some comments on the validity of the identification
(\ref{eq:ident}). Of course the real calculation of the term
$V_{eff}(q,\omega)$ or $M_q^{e}$, specially at finite temperature
would be more complex, and will certainly affect the quantitative
aspects of the identification. However we expect the
\emph{qualitative} features of the mode identification to be
robust. In particular the frequency of the acoustic plasmon will
of course be linear in $u\sim q$ at small $q$, and in the same way
the matrix element describing the coupling to the particles will
go as $M_q^{e}\sim q^{1/2}$. Note that this specific dependence of
the coupling constant is what makes the difference between the
coupling to this acoustic plasma mode and the coupling to normal
acoustic phonons. This will have of course consequences for the
temperature dependence of the resistivity that we will explore in
the next section. The most drastic approximation that we made was
to ignore the finite lifetime of the mode and to concentrate the
full spectral weight in a $\delta$-function peak. This
approximation is not essential, and the finite lifetime can of
course be taken into account. It would simply correspond to a
damping of the phonon mode. It simplifies however the subsequent
calculations and allows to extract the physics in a more
transparent way so we stick to it in the reminder of this paper.
On a quantitative level, both the broadening of the level $u'' q$
and the average energy $u' q$ are proportional to $q$, as
discussed in the previous section. For Bismuth a typical order of
the ratio $u''/u'$ is  $u''/u'\approx 0.25$ so we expect such an
approximation to be reasonably quantitative as well.

We can now use the standard diagrammatic analysis of
electron-phonon interaction, keeping in mind the difference in the
matrix elements, to obtain the various physical quantities when
the acoustic plasmon is playing a major role. In particular the
self energy (\ref{eq:selfphon}) is simply given, after summation
on the Matsubara frequencies by
\begin{multline}\label{eq:Sigma_res}
 \Sigma_e(q,i\nu_n) = \frac{1}{\Omega} \sum_{q} (M_q^e)^2 [\frac{b(\omega_q) + f(\xi(k+q))}{i\nu_n + \omega_q - \xi(k+q)}-\\
                       \frac{b(-\omega_q) + f(\xi(k+q))}{i\nu_n - \omega_q - \xi(k+q)}]
\end{multline}

Since we are interested in the low energy dissipation we
perform the analytical continuation $i\nu_n \rightarrow \nu +
\imath\delta_{+}$, then the limit $\nu\rightarrow 0$ and finally
extract the imaginary part of $\Sigma_e$. This last step leads to
a delta function constraints implementing the energy conservation
(see e.g. (\ref{eq:Econs})).
This equation gives us a 2D surface $\Omega_{0}$ of solutions $q_{0}$ for $q$.
However the above sum on $q$ in (\ref{eq:Sigma_res}) should be limited to
the values for which the acoustic plasmon exists. We should thus
only look at values of $q$ lower than $q^*$ which was defined in
(\ref{eq:upper}). For small angles only one of the masses dominates (\ref{eq:mass_comb}).
Combining this with (\ref{eq:velocity}) and
the fact that we work in $\omega\rightarrow 0$ limit, we deduce
that in a limit of small angles (nearby direction of high
symmetry) $c_{ac}(\theta)\sim \cos(\theta)$ which implies
$q_{0}(\theta)=q^{*}\cos(\theta)$. One has thus always $q_{0}(\theta)\leq q^{*}$.
Although potentially this criterion can depend
on the direction of $q$, based on above considerations, we can
assume here that the directions of $\vec{q}$ where $q^{*}(\theta)$
drops to extraordinarily small values are rare. In particular this
means that the surface $\Omega_{0}$ is well defined. To obtain
numerical responses we will take an isotropic criterion for the
upper cutoff.

\subsection{New $\rho(T)$ dependence}\label{sec:new_rho}

In the low temperature regime, the transport of the system, which is now dominated
by the presence of the acoustic plasmon, can thus be described by the effective
Hamiltonian (\ref{eq_hphonon}). One has of course to keep in mind that because of the
unusual $q$ dependence of the coupling to this bosonic degree of freedom (\ref{eq:ident}) the temperature
dependence of the resistivity will not be the one of the usual electron-phonon problem.

\begin{figure}
  \vspace{3cm}
  \includegraphics[width=0.6\columnwidth]{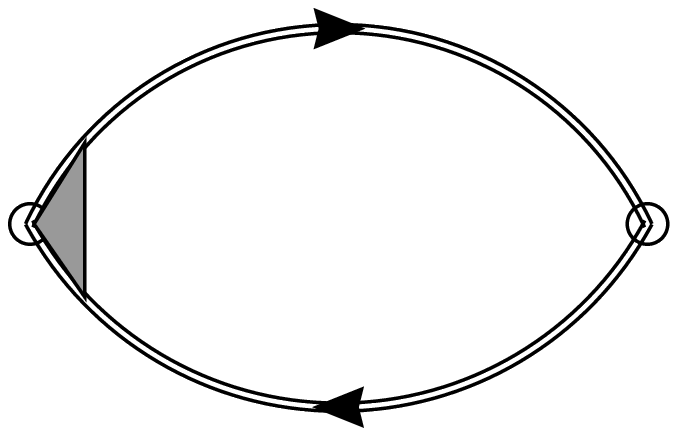}\\
  \includegraphics[width=0.99\columnwidth]{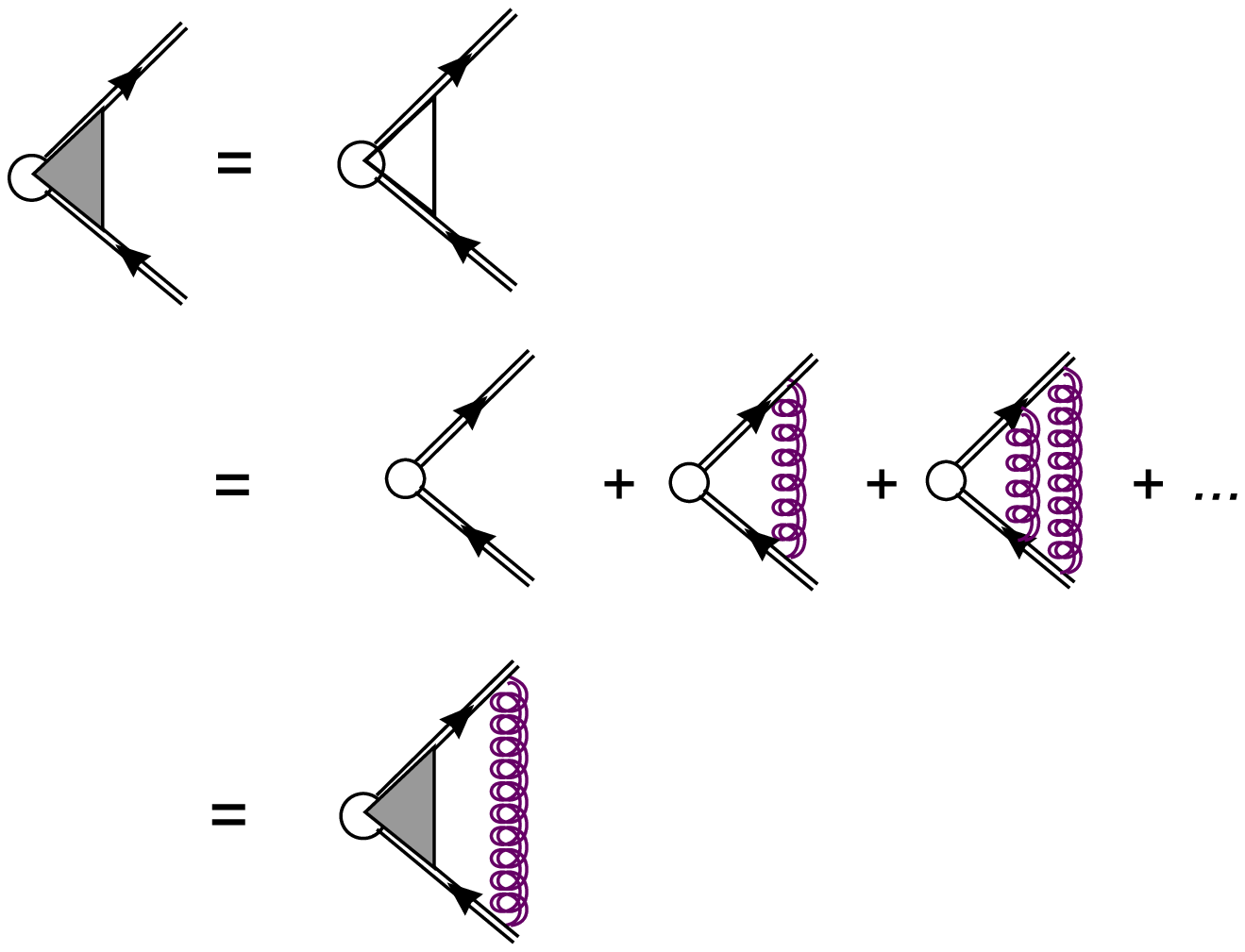}\\
  \caption{\label{fig:ladder_diag} a) The diagrammatic form of current-current correlation
  function. The double line is the full Green's function for the electrons or holes while the shaded box denotes
  the vertex correction which is shown in part b). In b) the ladder series for vertex correction is shown.
  The irreducible part consist on single plasmon exchange (purple coil line) between the fermions (electrons or holes).  As for normal phonons given the momentum dependence of the coupling
  constant $M_q$ and propagator, it is crucial to include the vertex correction to get the proper temperature dependence.}
\end{figure}

To compute the resistivity we use the Kubo formula and express the resistivity as the current-current correlation
function
\begin{multline}\label{eq:res_def}
    Re[\sigma_{xx}(\omega\rightarrow 0)]=\\
    lim_{\omega\rightarrow 0}\frac{ne^2}{\omega}Im[\langle j_{x}(Q=0,\omega)j_{x}(-Q=0,0)\rangle]
\end{multline}
We are interested in the uniform response, so we set $Q=0$, we
will omit $x$ index in the following. The current is simply given by
\begin{equation} \label{eq:currentdef}
 J = \sum_{k,\alpha} e_\alpha \frac{k_x}{m_{x\alpha}} c^\dagger_{k\alpha} c_{k\alpha}
\end{equation}
where $\alpha$ is the particle species, $e_\alpha$ and $m_{x\alpha}$ the corresponding charge and
mass in the direction of movement.
The corresponding diagram is shown on Fig.~\ref{fig:ladder_diag}a where the thick line and
shaded triangle indicate that both propagator and interaction
vertex are renormalized by $V_{eff}(q,\omega)$). The average in
(\ref{eq:res_def}) will be computed with the effective
hamiltonian of the problem given by (\ref{eq_hphonon}). It is important to note that given the topology
of the diagrams entering the vertex corrections there is indeed no double counting of diagrams if one replaces the
thick coil of Fig.~\ref{fig:ladder_diag} by the bosonic excitation as done in (\ref{eq_hphonon}).

The procedure is standard\cite{mahan_book}, and follows
closely the one for phonons. The vertex correction makes it
cumbersome, so we give here a derivation based on the memory
function that has the advantage to take directly the vertex
correction into account in a simpler way. The conductivity can be
expressed as
\begin{equation} \label{eq:basicmem}
 \sigma(\omega) = \frac{i\chi_0}{\omega + M(\omega)}
\end{equation}
where $\chi_0$ is the diamagnetic term, and  $M(\omega)$
is defined as:
\begin{equation}\label{eq:Mdef}
    M(\omega)=\frac{\langle F;F\rangle_{\omega}-\langle F;F\rangle_{0}}{-\chi(0)\omega}
\end{equation}
where the force $F$ is $F=[j,H]$, and the $\langle F;F \rangle_{\omega}$ denotes the
standard retarded correlation function at frequency $\omega$. The averages
are computed with the the part of the Hamiltonian that commutes with the
current. Given that $M=0$ if the current commutes with $H$ one recovers immediately
in that case from (\ref{eq:basicmem}) that the system is a perfect conductor.
In particular in the high temperature regime it shows immediately that the sole source of
resistivity is the electron-hole interaction and the difference of masses between the two species.

For low temperatures, in the case of the Hamiltonian (\ref{eq_hphonon}), and using the definition of the
current (\ref{eq:currentdef}) one obtains for $F$ (for simplicity we only kept two species
$e$ and $h$)
\begin{equation}
  F = \sum_{q,k,\alpha} e_\alpha M_{q\alpha} \frac{q_x}{m_\alpha} c^\dagger_{k+q,\alpha} c_{k,\alpha} [b^\dagger_{-q} + b_q]
\end{equation}
Each species gives thus a contribution to F-F correlation which is
shown on Fig.\, and for Matsubara frequencies equals to:
\begin{multline}\label{eq:M1}
    \langle F;F\rangle_{\omega_{n}} =
    \frac1{\beta^2\Omega^2} \sum_{\nu_1,\nu_2,k,q}\frac{M_q^2 q_x^2}{m^2} G(\nu_1,k)G(\nu_2,k+q)\\
    D(\omega_n + \nu_1 - \nu_2,q)
\end{multline}
\begin{figure}
  \includegraphics[width=0.9\columnwidth]{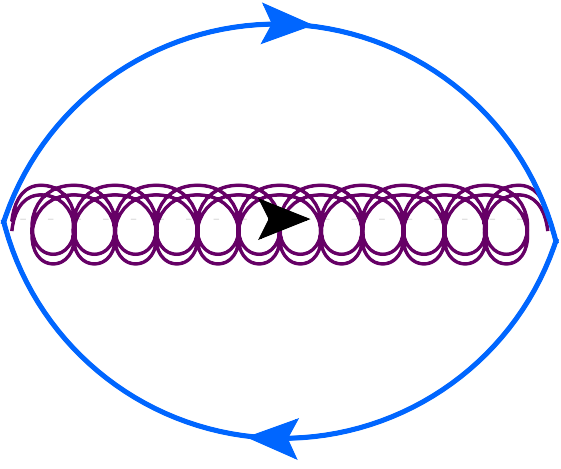}\\
  \caption{The force-force (F-F) correlation diagram. The solid blue line indicates propagator of fermion
  (one, for example slower, component of the Fermi liquid) while the purple coil is the plasmon propagator}\label{fig:F-Fcorr}
\end{figure}

The frequency summation can be performed which gives
\begin{multline}\label{eq:M2}
    \langle F;F\rangle_{\omega_{n}} =
    \frac1{\Omega^2} \sum_{k,q}\frac{M_q^2 q_x^2}{m^2}\\
    \frac{[f(\xi(k+q)) + b(-\omega_q)][f(\xi(k) - f(\xi(k+q) + \omega_q)]}{i\omega_n + \xi(k) - \xi(k+q)-\omega_q}\\
     - (\omega_q \to -\omega_q)
\end{multline}
After the analytic continuation one gets for the imaginary part of the function $M(\omega)$
\begin{multline} \label{eq:mfunc}
 {\rm Im} M(\omega\to0) = \frac{\pi}{\Omega^2} \sum_{k,q}\frac{M_q^2 q_x^2}{m^2} \frac{\partial f}{\partial \xi(k)} [f(\xi(k+q)) + b(-\omega_q)]\\
 \delta(\xi(k) - \xi(k+q) - \omega_q) - (\omega_q \to -\omega_q)
\end{multline}
We are interested in the temperature dependence of the
resistivity. We can thus for simplicity assume an averaged mass
over the Fermi surface, which would affect the prefactor but not
the temperature dependence. This allows to replace $q^2_x \to
q^2/3$. The term $\partial f/\partial\xi(k)$ constrains, at low
temperature, to have $k$ on the Fermi surface. One thus gets
\begin{equation} \label{eq:mfunc2}
\begin{split}
 {\rm Im} M(\omega\to0) & = \frac{-k_F}{6\pi\Omega} \sum_{q}\frac{M_q^2 q^2}{m} [f(\xi(k_F+q))+b(-\omega_q)]\\
         & \delta(\xi(k_F+q) + \omega_q) - (\omega_q \to -\omega_q) \\
         &= \frac{k_F}{6\pi\Omega} \sum_{q}\frac{M_q^2 q^2}{m} [f(\omega_q) + b(\omega_q)]\\
         &(\delta(\xi(k_F+q) + \omega_q) + \delta(\xi(k_F+q) - \omega_q))
\end{split}
\end{equation}
If the temperature is small the Fermi and Bose factors impose $q c_{\rm ac} \sim T$, and thus the argument of the $\delta$ function simplifies to
\begin{equation}\label{eq:Econs}
\delta(\xi(k_F+q) + \omega_q) = \frac{k_F}{m} q \cos(\theta) \pm
c_{\rm ac}(\theta) q
\end{equation}
leading to
\begin{equation} \label{eq:res_result}
 {\rm Im} M(\omega\to0) = \frac1{6\pi} \int_{\Omega_{0}} dq \frac{M_q^2 q^3}{2\pi^2} [f(\omega_q) + b(\omega_q)]
\end{equation}
where we applied the same reasoning as for the self-energy
(\ref{eq:Sigma_res}). The temperature can be rescaled out of the
above integral leading to
\begin{equation}\label{eq:res_result_final}
 {\rm Im} M(\omega\to0) = T^5 \int_0^{\beta q^*} dq \frac{M_q^2 q^3}{12\pi^3} [\tilde{f}(\omega_q) + \tilde{b}(\omega_q)]
\end{equation}
where the $\tilde{f}$ and $\tilde{b}$ and the Fermi and Bose
factors with $\beta=1$ and we took isotropic case. At low
temperature the integral tends to a constant, which can be
evaluated in the isotropic case, so the resistivity has a
\begin{equation}
 \rho(T) \propto T^5
\end{equation}
temperature dependence.

We thus see that the same electronic mechanism which at higher
temperature was giving the conventional Baber $T^2$ behavior will
smoothly lead to a $T^5$ behavior when acoustic plasmons begin to
govern screening. The new $T^{5}$ behavior comes essentially from
two approximations: the linear dispersion of bosons and $M_q \sim
\sqrt{q}$. The upper limit of the integrals in
(\ref{eq:res_result})- is the largest possible value of plasmon
momentum. This critical wavevector $q^*$, at which Landau damping
suppress the plasmon as a well defined particle, plays a similar
role than the Brillouin zone boundary for acoustic phonons. As we
discussed in the first section, one can also define a
corresponding temperature $T^{*}$ (see (\ref{eq:debplas}), which
is the analog of the Debye temperature and plays a similar role in
the resistivity.

If we take a zeroth order approximation for the interaction $M_q^{0}$
(see (\ref{eq:ident})) and $T^{*}\approx 1K$, we can
estimate the $A$ coefficient in the Deybe law (as defined in Eq.~1 of
Ref.~\onlinecite{Uher_Bi-rho77}). Along the trigonal axis we get
$A\approx 0.25[\mu\Omega cm]$ (in general
$A\in(0.25,0.6)[\mu\Omega cm]$) which is not far from the experimental
value $A\approx 0.15[\mu\Omega cm]$. The zeroth order
approximation is overestimated because $c_{ac}^{0}$ (and also
$c_{ac}^{RPA}$)is known to be overestimated, plus as we explained
while evaluating the integral (\ref{eq:res_result}) not all the 2D
surface of solution might be present.

\begin{figure}
  \includegraphics[width=0.9\columnwidth]{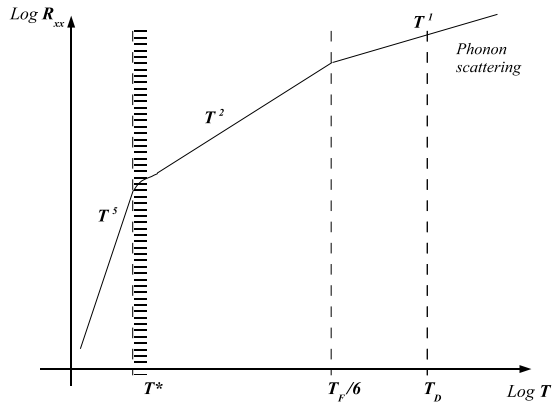}
  \caption{\label{fig:R(T)} The temperature dependence of resistivity in bismuth. Three regimes are distinguished $\sim T$, $\sim T^2$, $\sim T^5$. The lowest temperature regime is due to the presence of the collective acoustic plamon mode. The two high
  temperature regimes are due to the conventional Baber scattering mechanism.}
\end{figure}

A full temperature dependence of the resistivity is shown on
Fig.~\ref{fig:R(T)}.
Note that the measured $T^{*}$ will be always slightly smaller than the RPA one
$T^{*}_{RPA}$ due to many plasmon processes (which are not accounted in RPA
approximation) enhancing the Landau damping around $T^{*}$. In fact
one can expect that the RPA prediction $T^{*}_{RPA}$ works
better for polycrystals where acoustic plasmons have more decay
channels (for example they can scatter on surface waves)
alternative to Landau damping. This was indeed
observed experimentally\cite{Uher_Bi-rho77}.

Let us finally make a comment on the anisotropy of the
resistivity, to argue that the low temperature conductivity is not
strongly affected by a particular choice of $\vec{q}$ direction.
In order to estimate the effect we propose a zeroth order
approximation in which $\rho_{\vec{q}}\sim
(M_q^{0})_{\vec{q}}\langle b_{\vec{q}}^{\dag} b_{\vec{q}}\rangle$
where an average is taken at a given low temperature. The
direction dependence comes from the mass anisotropy and in this
respect we know that two factors behave in different ways.
\begin{itemize}
    \item from the reasoning given in Sec.~\ref{sec:acous_res}, in particular (\ref{eq:ident}),
    $$(M_q^{0})_{\vec{q}}\sim (c_{ac}^3)_{\vec{q}}$$
    while at the lowest order $c_{ac}^{0}\sim V_{Fs}\sim
    (m_{s\vec{q}})^{-1/2}$, which implies:
    $$(M_q^{0})_{\vec{q}}\sim (m_{s\vec{q}})^{-3/4}$$
    We expect this dependence to be stronger when we reach $q^{*}$.
    \item the average density of plasmons $\langle b_{\vec{q}}^{\dag} b_{\vec{q}}\rangle$
    at the lowest temperatures will be determined by the Taylor expansion of the Bose-Einstein
distribution $b(c_{ac}(\vec{q})q)$ which means $\langle
b_{\vec{q}}^{\dag} b_{\vec{q}}\rangle \sim m_{s\vec{q}}$ while at
the higher temperatures (around $T \approx T^{*}$) the density of
available slow fermions (necessary to build up a collective
excitation) sets the limit which is also $\sim m_{s\vec{q}}$.
\end{itemize}
Both contributions to $\rho_{\vec{q}}$ scale in an opposite way with
the mass thus we expect the overall direction dependence to be
rather weak (especially because a more refined approximation for
$\rho_{\vec{q}}$ would need to contain some average over the whole
Fermi surface).

We expect a stronger influence of anisotropy on $q^{*}$, because
this quantity depends only on the plasmon velocity
$u=c_{ac}-\imath\tau^{-1}$. The problem is rather complex and we
present a discussion of it in the
Appendix~\ref{ap:bismuth_plasmon}. Considering the anisotropy of
$q^{*}$ bring us to an approximation we have made while deriving
(\ref{eq:res_result_final}): we assumed that the plasmons
propagate equally well in all directions. As shown in
Appendix~\ref{ap:bismuth_plasmon} when more refined approximations
for the polarizability $\Pi(k,\omega)$ are used one gets more
isotropic $q^{*}$ values. Indeed the measured anisotropy of
$T^{*}$ is quite weak\cite{Uher_Bi-rho77}. This makes the
derivation of (\ref{eq:res_result_final}) self-consistent.

\subsection{Electron-plasmon coupling non-perturbative theory}\label{sec:plasmaron_intro}

As we saw in the previous sections, the presence of a collective mode has e drastic influence
on transport properties. The mapping to the effective Hamiltonian (\ref{eq_hphonon}) allows
to describe these effects. In the previous section we looked at the consequences of such a coupling
to the collective bosonic mode, in a lowest order calculation in the self energy.

However it is well known, for the case of electron-phonon coupling, that the combination
of the electrons and the phonon degrees of freedom can be carried out beyond this lowest
order and lead to a new composite quasi-particle, the polaron. For the case of an
acoustic plasmon, a similar phenomenon occurs\cite{Whitfield_polaron}.
In our case the corresponding
quasi-particle would be a plasmaron (electron accompanied by
plasmons). The creation operator of a plasmaron can be written
using a variant of the Lang-Firsov unitary transformation:
\begin{equation}\label{eq:plasmar_Lang-Firsov}
    \tilde{c_k}^{\dag}=c_{k}^{\dag}\exp(-\hat{S})
\end{equation}
where:
\begin{equation}\label{eq:plasmar_L-F_S}
    S=\sum_{q}\frac{\alpha_{q}}{e\hbar\omega_{q^{*}}} n_{q}(b^\dagger_q + b_{-q})
\end{equation}
where $\alpha_{q}\sim M_q/E_{F}$ measures the strength of the
boson-fermion coupling (in (\ref{eq:plasmar_L-F_S}) we assumed it
is real). The shift of fermion energy ($\sim Re[\Sigma_{boson}]$)
and mass ($\sim
\partial_k Re[\Sigma_{boson}]$) are known:
\begin{equation}\label{eq:plasmaron_Eshift}
    |\Delta E|=(\alpha+0.0123\alpha^2+0.00064\alpha^3+...)\omega_{q^{*}}
\end{equation}
\begin{equation}\label{eq:plasmaron_Mshift}
    \Delta m(\vec{q}) = (\alpha/6+0.024\alpha^2+...)m(\vec{q})
\end{equation}
where $m(\vec{q})$ indicates the mass in the $\vec{q}$ direction. The
above formulas were obtained in the so-called large plasmaron
picture. In the case of Bismuth the plasma is not rigid enough (contrarily to the
ionic lattice) to sustain self-localization, so a small plasmaron
picture is unphysical.

Naturally, when $\alpha\rightarrow 0$ all the above effects
disappears, but the question about the strength of full $M_q$ (not
$M_q^{0}$) is open and quite difficult to access theoretically. We
can only give an upper and lower limit for $\alpha$. First, $M_q$
must be smaller than the bare Coulomb interaction in order to make
the plasmon a well defined particle. The ratio of bare interaction
to kinetic energy is given by $r_{s}$ and as discussed in detail
in Sec.~\ref{sec:RPA_valid} $r_{s}\leq 1$ which implies
$\alpha<1$. The lower bound of $\alpha$ can be obtained
experimentally: recent optical spectroscopy
measurement\cite{Tediosi_plasmaron07} showed the existence of an
(optical) plasmaron in Bismuth which shows that the $\alpha$
coupling is sufficiently strong ($\alpha\sim 10^{-1}$) to give
observable effects.

The notion of plasmaron allows for some developments of
the theory beyond RPA. The numerical evaluation is left for future
investigations. We will come discuss some of these effects at
the end of Sec.~\ref{sec:RPA_valid}. However a brief inspection of
formulas (\ref{eq:plasmar_Lang-Firsov}-\ref{eq:plasmaron_Mshift})
has several consequences:
\begin{itemize}
    \item The dominance of the linear term in (\ref{eq:plasmaron_Eshift},\ref{eq:plasmaron_Mshift}) confirms the
accuracy of the first order approximation for electron-boson even
for $\alpha\sim 1$, which validates our perturbative approach in
$M_q$.
    \item A plasmaron remains a well defined fermionic particle, so the low energy model (\ref{eq_hphonon})
remains valid but with renormalized parameters. We can thus expect the $T^5$ behavior of the resistivity
to be still obeyed even in an intermediate coupling regime.
    \item from (\ref{eq:plasmaron_Mshift}) we see that in the presence of a plasmon cloud
    the mass of the heavier fermions further increases by $\Delta m_{s}$, which
    should give rise to a $q^*$ increase. More importantly the ratio
    of the geometric series which determines $\Delta m_{r}$ is
    proportional to $\sqrt{m_{r}/m_{r}}$ which means that the effect is the strongest in the direction where
    $q^*$ was smaller. Higher order effects thus tend to stabilize the acoustic plasmon
    \item by definition (\ref{eq:plasmar_L-F_S}) a fermion is
    mostly accompanied by plasmons propagating in the same direction ($\vec{q}_{ferm}\parallel
    \vec{q}_{boson}$); these plasmons are also mediating the effective
    interaction between electron and holes and thus the part of
    interaction $\parallel \vec{q}_{ferm}$ is enhanced. This
    goes towards the unidirectional scenario advocated in Appendix~\ref{ap:bismuth_plasmon}.
\end{itemize}

\section{Optical spectroscopy: magneto-plasmon} \label{sec:magnetop}

An obvious question is whether this new quasi-particle,
is responsible for the lowest temperature resistivity
data, can be probed by other measurements.

Let us consider the consequences of the existence of the plasmon
mode when there is a magnetic field acting on the system. We
assume that the plasmon propagation direction is perpendicular to
the external magnetic field, the geometry which is usually used in
optical spectroscopy. We also assume that in this plane electrons
are heavier than holes (this can change depending the bismuth
sample orientation). As we will the low energy collective
excitation can be probed by measuring the optical conductivity of
the system.

\subsection{Dispersion relation}

The inconvenience of studying the acoustic plasmons with optical
methods comes from the fact that in the limit $q\rightarrow 0$,
which is the resonance condition with photons, by definition we
have $\omega \rightarrow 0$. The way to overcome this difficulty
is to introduce a magnetic field. In the case of a standard
(optical) plasmon the frequency of collective excitation
(propagating perpendicular to the magnetic field) is increasing by
the magnetic field as $\omega=\sqrt{\omega_{p}^2+\omega_{c}^2}$.
This phenomenon can be explained intuitively: the movement of a
charge under a magnetic field requires an extra kinetic energy (to
overcome the magnetic field vector potential. This result was
found in several different ways: using RPA polarizability,
equation of motion technique or from simple hydrodynamic picture
of fluctuations. The reason why such a remarkably simple relation
holds comes from Kohn's theorem: in a translationally invariant
system, for $q\rightarrow 0$, the magnetic field does not affect
the inter-electron interactions encoded in $\omega_{p}$. This
quite general theorem holds also for multi-component plasmas.

Indeed for the two component plasma problem under magnetic field
the following, a relation, valid for any value of the magnetic field, in the $q\rightarrow 0$ limit was found\cite{Noto_magnetoplasm74}:
\begin{multline}\label{eq:omega_ac}
    \omega_{\pm}^{2}=\frac{1}{2}[(\omega_{p}^{e2}+\omega_{p}^{h2}+\omega_{c}^{e2}+\omega_{c}^{h2})\pm\\
    \sqrt{(-\omega_{p}^{e2}+\omega_{p}^{h2}-\omega_{c}^{e2}+\omega_{c}^{h2})^2 +
    4\omega_{p}^{e2}\omega_{p}^{h2}}]
\end{multline}
When $|\omega_{c}^{e}-\omega_{c}^{h}|$ is small we find the
intuitive relations $\omega_{+}^2\approx
\omega_{p}^{e2}+\omega_{c}^{e2}+\omega_{p}^{h2}+\omega_{c}^{h2}$
and $\omega_{-}^2\approx |\omega_{c}^{e}\omega_{c}^{h}|+O(q)$,
which allows for a clear identification of optical and acoustic
plasmon when $B\rightarrow 0$. We see that the acoustic mode
develops a gap, which should be observable in optics as a second,
lower frequency plasmon edge. In the limit when
$|\omega_{c}^{e}-\omega_{c}^{h}|\gg
|\omega_{p}^{e}|,|\omega_{p}^{h}|$ the $\omega_{\pm}$ modes in
(\ref{eq:omega_ac}) become hole and electron like respectively. It
is relatively simple to reach this limit (when the second term
under square root is small) in bismuth. The reason is that we
always have $m_{r}\ll m_{s}$, which implies that
$|\omega_{c}^{e}-\omega_{c}^{h}|\sim
|\omega_{c}^{e}|,|\omega_{c}^{h}|$. We thus expect $\omega_{-}$ to
have an electron-like (in general heavier particle-like) magnetic
field dependence.

From (\ref{eq:omega_ac}) we can deduce how the frequency of
the lower plasmon edge evolves with magnetic field. The temperature
dependence of $\omega_{\pm}$ is also frequently studied in optical
experiments. The temperature dependence enters
(\ref{eq:omega_ac}) via $\omega_{p}^{e,h}\sim n_{e,h}$. The
density of carriers $n_{e,h}$ can change with temperature due to
thermal excitation from the valence band. Because of the very small gap
at the bottom of electron pockets $\Delta\approx 13.7 meV$ (see Fig.~\ref{fig:band-structure})
this effect is particularly important for bismuth. Then a second type of
holes, as light as electrons, will emerge, and due to the global
compensation of electrons and holes $n_{e}=n_{h}+n_{lh}$ the total
number of carriers will increase. In the low field limit only the
standard (high energy) optical plasmon frequency $\omega_{+}(T)$
will be affected.

In the high field limit the $\omega_{-}$ will acquire a rather
strong temperature dependence since $\omega_{-}^2(T)\approx
(\omega_{p}^{e}(T))^2+(\omega_{c}^{e})^2$ (assuming that the
electrons are heavier). From the charge neutrality condition the
approximate formula for $\omega_{p}^{e}(T)$ can be
given\cite{Dirkgroup}:
\begin{equation}\label{eq:wpe(T)}
   (\omega_{p}^{e})^2=\frac{4e^2}{\sqrt{2}\pi\hbar}\frac{\bar{m}_{e}^{3/2}}{m_{x}^{e}}Li_{1/2}(\exp(\beta(-E_{Fe}-\mu)))(k_{B}T)^{3/2}
\end{equation}
where $e$ is the electron charge, $\bar{m}_{e}$ is geometrical
average of electron masses and $m_{x}^{e}$ is a mass in a given
direction (all masses in electron mass unit).
$Li_{1/2}(\exp(\beta(-E_{Fe}-\mu)))$ is the value of the
incomplete polylogarithm function of order $1/2$:
\begin{equation}\label{eq:Li}
   Li_{1/2}(a)=\int_{0}^{\infty}dx\frac{\sqrt{x}}{a\exp(x)+1}
\end{equation}
This gives the temperature dependence of the $\omega_{-}$ for
a very large magnetic field when this mode merges with the electron mode
(in general the heavier carrier mode).

\subsection{Longitudinal f-sum rule}\label{ssec:frule}

The third quantity accessible from optical measurements is the
relative intensity of two plasmon edges. This can be studied
quantitatively using the longitudinal f-sum rule.

In a one component plasma it can be proven that the optical
plasmon is the only collective excitation as $q\rightarrow 0$. It
comes from an exact sum rule (see p.358 in
Ref.\onlinecite{mahan_book}):
\begin{equation}\label{eq:f-sum}
    \int_{0}^{\infty}d\omega\omega
    Im(\epsilon(q,\omega)^{-1})=-\frac{\pi}{2}\omega_{p}^{2}
\end{equation}
An analog sum rule was derived for a multi-component plasma (see
for example Ref.~\onlinecite{PhysRevE.62.5648}). It was also
shown\cite{PhysRevB.53.9797} that in the presence of a non-local
(q-dependent) potential the right hand side of (\ref{eq:f-sum})
can change. This is particularly important for a two component
plasma when the non-local part might differentiate between intra-
and inter-component interactions (defined as $V_{\rm
eff}^{\alpha\beta}(q)$ in (\ref{eq:collecmode}) \footnote{strictly
speaking in Bi this is the case for $q\neq 0$, but arbitrarily
small}). Then the sum rule is not completely exhausted by the
$\omega_{+}$ mode and there is some spectral weight left for the
$\omega_{-}$ plasmon \footnote{see also
Ref.~{\protect\onlinecite{Das_Sarma_ac-plasm99}}, where the acoustic plasmon
intensity increases for $d\neq 0$ a condition which in their case
implies $V_{11}(q)\neq V_{12}(q)$}. These findings prove that the
finite spectral weight of acoustic plasmon does not violate the
longitudinal f-sum rule.

One can quantify this last statement by computing the strength of
each plasmon, which in fact corresponds to $Im[\Sigma_{e}(q)]$
evaluated in Sec.~\ref{sec:acous_res}:
\begin{equation}\label{eq:spectr_weight}
    W_{\pm}(q)=\frac{\pi}{|\partial Re[\epsilon(q,\omega)]/\partial\omega|_{\omega=\omega_{\pm}}}
\end{equation}
where $\epsilon(q,\omega)$ is a dielectric function defined as the
denominator of Eq.~\ref{eq:collecmode}: $\epsilon(q,\omega)=1 -
V^{ee}_{\rm coul}(q)[\Pi_{hh}(q,\omega) + \Pi_{ee}(q,\omega)$.
Using the formulas given in Sec.~\ref{sec:mecha} and the Lindhard
approximation for polarizability, we find (in the $q\rightarrow 0$
limit):
\begin{equation}\label{eq:epsil_derivative}
    \partial Re[\epsilon(q\rightarrow 0,\omega)]/\partial\omega|_{\omega=\omega_{-}}\approx
    (r_{v}-R)\ln(\frac{\omega_{-}-V_{Fs}q}{\omega_{-}+V_{Fs}q})
\end{equation}
where $r_v$ and $R$ are characteristic, constant parameters of the two
component plasma (they are of order $1$, for precise definitions
see Appendix~\ref{ap:bismuth_plasmon}). Taking into account $\lim_{B \to
0}\omega_{-}=c_{ac}q$ (neglecting $Im[u]$) we deduce that the
numerator of the logarithm limits $W_{-}(q)$:
\begin{equation}\label{eq:spectr_weight-r}
    \lim_{B\rightarrow 0} W_{-}(q)\sim [(r_{v}-R)\ln((c_{ac}-V_{Fs})q)]^{-1}
\end{equation}
From (\ref{eq:spectr_weight-r}) one recovers the
Landau damping formula: when $c_{ac}\rightarrow V_{Fs}$,
$W_{-}\rightarrow 0$. The larger $c_{ac}$ is with respect to
$V_{Fs}$, the larger spectral weight the acoustic plasmon will
contain. From (\ref{eq:spectr_weight-r}) we also see that, when
$c_{ac}\neq V_{Fs}$, the $W_{-}(q)$ is an increasing function of
$q$; it is difficult to expand this formula in $q\rightarrow 0$
limit, but using l'Hospital rules (for the derivative of $\ln(q)$)
one can confirm our argument given in Sec.\ref{sec:acous_res} that
$V_{0}^{ac}\sim q$.

The other implication of (\ref{eq:spectr_weight}) is that
inducing any gap in $\omega_{-}(q=0)$ spectrum will increase
$W_{-}(q=0)$. This is precisely what magnetic field does as shown
in (\ref{eq:omega_ac}). In particular in a strong magnetic field
$\omega_{-}\rightarrow \omega_{c}^{e}+\omega_{p}^{e}$ which means
that the acoustic mode becomes a mode of heavier (electron in our
example) plasma component. Then, there must be also a
redistribution of the relative spectral weights of the two plasmon
edges $\omega_{\pm}$; the $\omega_{-}$ mode will be more and more
pronounced. To be more precise in the equation
(\ref{eq:spectr_weight}), to get the intensity of $\omega_{-}$
mode, we substitute $\epsilon(q,\omega)$ of the heavier component
plasma. The $Re[\epsilon(q\rightarrow 0,\omega_{-}(q))]\rightarrow
q^2$ so the denominator of (\ref{eq:spectr_weight}) gradually
approaches zero and does it in the same way than
$Re[\epsilon(q\rightarrow 0,\omega_{+}(q))]$. This resembles the
fact that the spectral weight $W_{-}(q\rightarrow 0)$ must become
as significant as $W_{+}(q\rightarrow 0)$. As the magnetic field
increases $W_{-}(q\rightarrow 0)$ gradually acquires all the density
fluctuations of the heavier carriers.

\subsection{Summary of the expected effects}

The findings of two previous sections are summarized in
Tab.~\ref{tab:mag}.
\begin{figure}
\begin{tabular}{|c|c|c|}
  \hline
  property & low B & large B \\
  \hline
  $\omega_{-}(B)$ & $\sim\sqrt{|\omega_{c}^{e}\omega_{c}^{h}|}$ & $\sim\sqrt{(\omega_{c}^{e})^2+(\omega_{p}^{e})^2}$ \\
  $\omega_{-}(T)$ & T independent & $\sim \omega_{p}^{e}(T)$ \\
  $W_{-}(B)=1/\epsilon(\omega_{-})$ & $W_{+} \gg W_{-}\sim B$ & $W_{-}\rightarrow W_{+}$  \\
  \hline
\end{tabular}
\caption{\label{tab:mag} Summary of the expected field dependence
of the two plasmon edges. We present the magnetic and temperature
dependencies of the lower energy plasmon edge and the magnetic
field dependence of its amplitude. The low and high field regimes
are distinguished}
\end{figure}
In order to define the border between \emph{low B} and
\emph{large B} we define the ultraquantum (UQ) limit for which the
magnetic field freezes the motion of carriers in a plane
perpendicular to it. Then $\omega_{c}$ dominates over
$\omega_{p}$. Due to the particular band-structure of bismuth and
the field dependence of the chemical potential electrons and holes enter
to UQ regime at nearly the same field ($\approx 10T$). Thus the
above mentioned condition $|\omega_{c}^{e}-\omega_{c}^{h}|\ll
\omega_{p}$ is valid up to quite high fields. The limit of validity
of this condition is for fields smaller but of the order of the ones
of the UQ regime. The value will change depending whether one works with pure
Bi or $Bi_{1-x}Sb_{x}$ (x<0.07) where the UQ limit is reduced.

There are effects not included in the above analysis. First we
observe that, because of charge neutrality requirement, and the small
value of the gap below the electron pocket, the chemical potential $\mu$ does depend on
the magnetic field. Thus we expect:
\begin{equation}\label{eq:nu(B)}
   \mu(B) \Rightarrow n(B) \Rightarrow \Delta\omega_{p}^{e,h}(B)
\end{equation}
but this extra dependence is significant only for
fields above the UQ regime. In the same regime one
may also expect an anomalous Zeeman splitting and large
magnetostriction which in principle can change the band parameters
(masses). A large spin-orbit coupling in Bismuth is responsible
for these effects. They are beyond the scope of this
paper, which is dedicated rather to low field effects, and we
leave them for future investigations.

The other source of finite gap in the spectrum of acoustic plasmon
at $q=0$ is the finite tunneling probability $\Upsilon_{\perp}$
between two types of carriers (electrons and holes in our case).
According to Ref.~\onlinecite{Das_Sarma_ac-plasm99} the additional
gap $\Delta_{SAS}$ (using the notations of this paper) will be proportional to
tunneling (or scattering) probability (see Eq.~26 in
Ref.~\onlinecite{Das_Sarma_ac-plasm99})
$\Delta_{SAS}\sim\sqrt{V(q=0)\Upsilon_{\perp}}$.

Due to Bi band structure the energy-momentum conservation strictly
forbids such a scattering, thus at $T=0\Rightarrow
\Upsilon_{\perp}=0\Rightarrow \Delta_{SAS}=0$. The situation could
be slightly more complicated at finite temperature for which we can
find a combination of energy conserving states, but the
conservation of $\vec{k}$ is still never fulfilled (electron and
hole pockets are very far in momentum space). We conclude that
this type recombination processes, even in the higher order
scattering events, should not affect the zeroth order prediction for
$\omega_{-}(B\rightarrow 0,T)\rightarrow 0$ given before.

\section{Discussion} \label{sec:discuss}

In this section we want to discuss the limits of validity of
our theory and the physics that can be expected when we reach those limit.
The experimental relevance of these effects will be presented.

\subsection{Validity of RPA approximation} \label{sec:RPA_valid}

In order to assert the validity of analysis of the previous sections,
which is mostly based on an RPA approximation of the interaction terms,
it is important to estimate first the strength of the interactions in
the case of Bismuth. At ambient pressure the ratio $r_{s}$ of the potential
energy to kinetic energy are respectively $r_{s}^{e}\approx 0.2$ and $r_{s}^{h}\approx 1.5$
for the electrons and the holes. These rather low values are due mostly to
the very high, high frequency, dielectric constant of the background
$\varepsilon_{\infty}=88$. It is also important to note that a part of the
Coulomb interaction, namely the Hatree-Fock terms with exchange interaction
$V_{\rm Coul}(q)/\epsilon_{TF}(q)$ with Thomas-Fermi screening
was already taken into account during a self-consistent band
structure calculations, leading to renormalized dispersions
$\varepsilon_{e,h}(k)$\cite{Allen_Bi-band95}.

Although this question is difficult to address quantitatively it is useful
to estimate if for the particular case of Bismuth we could trust the RPA approximation.
Usually the following arguments are used to justify the RPA re-summation shown on
Fig.~\ref{fig:electr_self-energy} are: i) the small number of carriers implies that the long range
character of interactions plays a major role ($V(q)$ large for $q\rightarrow 0$).  Then
the diagrams with the largest divergence number (RPA series) are the most
important. This is the case in bismuth despite the simultaneous validity of the ``dense plasma'' regime $r_{s}<1$;
ii) we are primarily interested in the limit of long wavelength excitations (below $q^*$) thus the influence of local field
effects is moderate.

In addition to these two arguments that support the RPA
approximation, the particular combination of band-structure
parameters of bismuth contributes to a further suppression of the
diagrams outside the RPA series. Indeed in Bismuth the Fermi
surface consists of distant pockets for electrons and for holes. A
significant part of the exchange diagrams (the inter-pocket ones)
requires quite large momentum exchange, so in the limit of small
momenta, as discussed above, ($V_{\rm
eff}(q\rightarrow\infty)\rightarrow 0$) their contribution must be
very small. In addition such processes are irrelevant in the RG
sense \cite{Abrikosov_BiRG73}. Another contributing factors comes
from the fact that electrons along the trigonal axis can be
considered as Dirac particles, while holes are very light in the
perpendicular plane. If one take a (one component) gas of spinless
fermions with linear dispersion then bubble approximation (each
bubble with only two external interaction lines) is exact. In
particular, in one dimension, for such a Dirac spectrum, the RPA
approximation would be exact. In higher dimensions one can thus
expect that the amplitude of non-bubble diagrams should be at
least be reduced when $m\rightarrow 0$. Finally both the large
mass anisotropy (on the surface of each pocket) and the strong
spin-orbit coupling (lowering the orbital momentum $\hat{j_{e}}$
quantum number ($\hat{j_{e}}=\hat{l_{e}}+\hat{s_{e}}$) contribute
to the weakening of the electron-electron exchange processes.
Although none of this arguments is of course rigorous or final,
they suggests that the RPA approximation in Bismuth is indeed a
very good starting point to understand the properties of this
material.

Improving above the RPA is of course a very difficult proposal.
Among the diagrams one would have to consider there are multi
acoustic plasmon processes (optical plasmon has a high energy so
we can safely exclude it). Some of them are clearly suppressed.
First, $M_{q}=0$ which implies that tadpole-like diagrams mediated
by acoustic plasmons gives no contribution. Second, a
multi-plasmon scattering processes has a natural cut-off $q^{*}$
above which they are strongly Landau damped. This mechanism of
$q_{\rm eff}^{*}$ reduction was invoked in the context of the
Fig.~\ref{fig:R(T)}. In fact given the acoustic nature of
plasmon excitation one might be tempted to introduce a variant of
the Migdal theorem in order to exclude processes with crossed
plasmon lines. The Migdal theorem in our case has the following
form: if we compare the available momentum phase space we find
that n-crossed multi-plasmon lines will generate corrections
proportional to $\lambda^{n}\omega_{q}/E_{F}$, where
$\lambda\simeq M_{q}$. In order to estimate the value of the
interaction we note that the electron plasmon coupling cannot be
larger than the bare Coulomb interactions $\lambda < e^2 k_{F}\sim
E_{F}$. Similarly the energy transferred by the acoustic plasmons
is of the order of $\omega_{q}\leq T^{*}<E_{F}$. Given the
hierarchy of energies between $T^{*}$ and $E_F$ we see that even
without the large mass ratio that exists in the case of the
electron-phonon coupling, here we have a justification of a
Migdal-like theorem to drop the higher order crossed plasmon
diagrams.

Of course, even after excluding all the above mentioned processes we are
still left with many diagrams, which can be constructed around a
dominating RPA series. The first class of multi plasmon diagrams
which are left are the rainbow ones. Usually diagrams of this type
are responsible for renormalization of particles bare dispersion,
which we have accounted during the discussion of the plasmaron in
Sec.~\ref{sec:plasmaron_intro}. The second class of diagrams are
the vertex corrections inside a single electronic bubble. The
third class which would make the problem highly non trivial
corresponds to interactions between several electronic bubbles
which would lead to an interaction between the plasmon modes or
interactions among several bubbles of the ladder series. How to
take into account all these terms is of course going well beyond
the scope of the present study. Based on various arguments one can
expect on a phenomenological level a renormalization of the
plasmon velocity. A phenomenological model that could account for
those extra contribution would be to add an interaction term
between the plasmons in our low energy Hamiltonian
(\ref{eq_hphonon}) of the form
\begin{equation} \label{eq:phenint}
 H_{\rm pl-int} = \frac1\Omega \sum_{k_1,k_2,q} g(q) b^\dagger_{k_1 + q} b^\dagger_{k_2 - q} b_{k_1} b_{k_2}
\end{equation}
the interaction between plasmon term $g(q)$ can in principle be estimated from the above mentioned diagrams,
and this is left for a a future study. This phenomenological term could thus open the route to tackle the situation
$r_{s}>1$.

\subsection{Pressure induced semimetal decay}\label{sec:pressure}

One way to control the physics of the acoustic plasmon could be to
consider the effect of pressure on these materials. Indeed the
tiny Fermi surface in Bismuth is quite fragile versus Sb doping or
pressure -- the pockets empties and the material becomes a
semiconductor. These dependencies are well established both on
theoretical (DFT pseudo-potentials method in
Ref.~\onlinecite{Norin_Bi(p)DFT77}) and experimental
\cite{Balla_Bi(p)65, Mendez_Bi(p)81} side. Obviously the
collective excitations such as the plasmon will be affected by
such a semimetal-semiconductor (SM-SC) transition.

At the first naive level the change of carrier density affects
only optical plasmon. The velocity $c_{\rm ac}$ (see
Appendix~\ref{ap:bismuth_plasmon}) depends on the ratio of masses
and Fermi velocities of the two types of carriers. These crucial
parameters $m_{e,h}$ or $V_{F}^{e'h}$ seem to be constant up to
very low carrier concentrations \cite{Norin_Bi(p)DFT77}. However
when $n\rightarrow 0$, the screening of the interaction is
drastically affected and $r_{s}\rightarrow\infty$. The corrections
beyond RPA, discussed in the previous section, start to be
important. On the other hand we know that, at least for Sb doping,
the gap (at L points) is being closed, thus $\varepsilon_{\infty}$
increases, extending the validity of the ``high density plasma''
regime. It means that it should be still possible to define
plasmons as quasiparticles of the system and the physics which we
described in the previous sections should still be applicable.

Non-trivial effects may arise only near the  critical doping
$\delta_{c}$ or pressure $p_{c}$ for which $r_{s}\rightarrow\infty$. Two elements will play a role:
i) as described in the previous section (and at the end of
Sec.~\ref{sec:acous_res}) the most important effect emerging with an increasing $r_{s}$ is the appearance of strong
plasmon-plasmon interactions. The interaction term (\ref{eq:phenint}) could potentially lead to an instability
with a new minimum of the energy at a finite $q$. In that case the corresponding energy gain would favor
a semimetal (correlated liquid) versus a semiconductor; ii)
contrarily  to $\omega_{ac}(q)=c_{ac} q$, the frequency of the optical plasmon $\omega_{p}\sim n$ is
decreasing when $n\rightarrow 0$. At some point the two collective excitations will merge. This should change both their
dispersion as well as the physics of the system.

On the experimental side, the regime we are discussing above was
recently studied by means of optical
spectroscopy\cite{Tediosi_pressure}. A transfer of the spectral
weight to the plasmaron peak was observed. This implies that, as we
expect, plasmons are able to survive (or can be even enhanced). In
these experiments a deviation from the Fermi liquid theory was found,
with an abnormal rigidity of the metallic phase near $p_{c}$. What
was observed can be interpreted as an abnormal increase of the collective mode
mode frequency (deviation from a single mode RPA at the lowest
temperatures). The extension of the theory proposed in the
previous section, (\ref{eq:phenint}) can perhaps be used in such a regime
to explain these effects.

\subsection{Comparison with other multi-valley semimetals}

As already mentioned in the introduction, a few other
examples of the multi-valley semimetals are known, with a band
structure similar to Bismuth. These other systems are thus potentially described
by a similar theory than the one that we have introduced in this paper.

The most intensively studied material in this category is
graphite. In this case electrons and hole pockets are placed
alongside, and are quite similar in shape, thus the carriers mass
(and velocities) ratio is close to $\simeq 1$.
In that case (see Appendix~\ref{ap:bismuth_plasmon})
$q^{*}$ shall be very small and the acoustic plasmons are always
overdamped. We thus expect a standard Baber $T^2$ resistivity down
to the lowest temperatures. This is indeed what is observed
experimentally\cite{Uher_graphite84}.

The second material, recently investigated, is $1T-TiSe_{2}$. The
band structure resembles strongly the one of Bismuth. It consist
of three electron and one hole pocket with large mass
differences\cite{Aebi_EI07}, $m_{h}=0.23$,
$m_{ex}=5.5$,$m_{ey}=2.2$ and $\epsilon_{\infty}=44$ (due to the
large $Se$ polarizability)\cite{Wilson_TiSe-diel78}. This suggests
that we can have another family of materials where acoustic
plasmons play a major role. The band structure gives a finite
amount of acoustic plasmons in the system ($k^{*}\neq 0$) but also
a quite large $r_{s}$, which resembles more the situation
described in Sec.~\ref{sec:pressure} than the one of pure bismuth.
This is most likely the reason why experimentally the low energy
physics of $1T-TiSe_{2}$ is different than in Bi. A density wave
phase transition was measured at $T_{DW}=63K$ and the related
reconstruction of Fermi surface was revealed by ARPES measurements
\cite{Aebi_EI07}. Interestingly the unusual increase of
resistivity \cite{Levy_rhoTiSe79} (attributed to an increase of
the electron-hole scattering rate at the transition) just above
the $T_{DW}$ seems to be similar to what was observed in Bi very
close to the critical pressure $p_{c}$. The formation of an
excitonic liquid was suggested to explain the transition at
$T_{DW}$. Whether one can make contact between such an excitonic
liquid description and the interacting plasmon theory that we
described in the previous section is a challenging theoretical
question that we leave for future studies. Note that at the lowest
temperature in $1T-TiSe_{2}$ superconductivity was recently
observed \cite{Tutis_SC09} suggesting additional lines of studies
that would use the similarity between the plasmons and the
acoustic phonons as an exchange particle.

\section{Conclusion} \label{sec:conclusion}

We have presented in this paper a theory of transport in
semimetals, concentrating specially on the case of Bismuth. We
have shown that the physical properties of these systems are
dominated by the presence of an acoustic plasmon mode at low
temperatures. This mode that we derived in an RPA approximation of
the interactions lead to a drastic change of the transport
properties, in particular their temperature dependence, compared
to the standard Baber mechanism which is normally invoked for such
materials. We showed in particular that it would lead to a $T^5$
behavior of the resistivity below a certain energy scale $T^{*}$
dependent on the interactions that we computed. Above this energy
scale a normal $T^2$ like behavior is recovered for the
resistivity. Our results are in agreement with the observed
resistivity in Bismuth. We examined several other consequences of
the existence of such a mode and showed that it would lead to a
double plasma edge in optical magnetotransport experiments. Recent
measurements of the optical conductivity agree well with our
predictions.

The main contribution of our paper is thus to show the importance
of such a plasmon mode in these systems. This opens the way to many
lines of work centered around the existence and role of such
collective modes. First it is important to explore the
consequences of these collective modes for other transport
properties. One of the most important is of course the Nernst
effect which in Bismuth is one of the largest reported. Analysis
on how the plasmon modes can modify the Nernst transport is a non
trivial questions that we plan to analyze in details. Another
important direction is to go beyond the RPA analysis. This is
necessary to investigate other semimetals such as Sn-doped Bismuth
or $1T-TiSe_{2}$ or to take into account the effects of pressure
on Bismuth. In that cases, the interactions between the plasmons
play a more important role. We have suggested a phenomenological
model which can potentially be used to tackle these effects.
However its analysis is highly non trivial and this will provide
certainly an exciting line of investigations for the future.

\appendix

\section{Acoustic plasmons in bismuth} \label{ap:bismuth_plasmon}

We give here some arguments and energy scales for the acoustic plasmons
in Bismuth.

The existence of acoustic plasmons was initially investigated for
Bismuth \cite{Bennacer_ac-plasm-Bi89} by means of a simple
spherical Fermi surface model with $m_{s}=m_{h}=0.14$ and
$m_{r}=m_{e}=0.047$. Based on this approximation it was concluded
that the acoustic plasmon was very weak ($k_{c}=0.02k_{Fs}$),
although the authors also mentioned in their conclusion the issue
of Fermi surface elipticity.

Solving when taking into account the full elipticity is a
formidable, albeit necessary task, as is obvious when looking at
the parameters given in Table~\ref{tab:masses}. The dispersion
relation of the collective mode is given by (\ref{eq:collecmode})
in which the full masses should be inserted in the $\Pi(q,\omega)$
functions. Since the dispersion is quadratic one can in principle
rescale the integration over $k$ in (\ref{eq:lindsum}). One thus
has
\begin{multline}\label{eq:gen_polar}
 \Pi^0_{\alpha\alpha}(q,\omega) =\\
 (m_{\alpha1} m_{\alpha2} m_{\alpha3})^{1/2}\tilde{\Pi}^0(q_1/\sqrt{m_{\alpha1}},q_2/\sqrt{m_{\alpha1}},q_3/\sqrt{m_{\alpha3}},\omega)
\end{multline}
where $\tilde{\Pi^0}$ is the normal Lindhard function with all the masses set to $1$.
For the case $\omega/\tilde{q} = c$ constant and $\tilde{q} \to 0$ the Lindhard function becomes particularly simple
\begin{equation}
 \tilde{\Pi}^0(\tilde{q}\to 0,\omega=c \tilde{q}) = \frac{\tilde{k}_F}{\pi^2} \left[-1 + \frac{s}{2}\log\left|\frac{1+s}{1-s}\right|\right] -\imath \frac{\tilde{k}_F s}{2\pi} \theta(1-s)
\end{equation}
where $\tilde{k}_F^2/2 =E_F$ the Fermi energy of the corresponding
species. Note that the two Fermi energies are not necessarily the
same for the two species of particles and must be determined by
the neutrality condition. When going to higher order diagrams
(vertex corrections) one realizes that not all scattering
directions are equally probable. This means that the coefficient
in front of $\tilde{\Pi}$ in Eq.\ref{eq:gen_polar} will change,
because it depends on the anisotropic mass tensor. However the
form of the Lindhard function should not be affected by this
procedure, thus the functional dependence of plasmon's dispersion
relations should remain unchanged.

One can thus try to simplify such an equation. An extension of RPA
(eRPA) taking into account the ellipsoidal
shape\cite{Bennacer_anisotrop92} made the approximation to compute
$c_{\rm ac}$ of averaging the faster component on a spherical
Fermi surface, while retaining the ellipsoidal shape of the
heavier (slower) component. If one follows this procedure, and for
example on the trigonal axis the electrons are faster than holes,
so we need to substitute $m_{h}=m_{h3}=0.63$ which is already
significantly larger. In the direction perpendicular to trigonal
the situation is even more complex. Holes are quite heavy along
the trigonal axis direction which gives large spherical mass and
suggests that they are always the slower component. However they
are very light in the perpendicular direction (see
Tab.~\ref{tab:masses}). It is then easy to forget that electrons,
while extremely light along trigonal axis, are (even in average)
significantly heavier than holes in any perpendicular direction. A
spherical approximation (even the eRPA) would thus need to assumes
to be able to do the average an opposite mass order than it is in
reality! This suggest that a different approach is necessary.

A simple average over a spherical Fermi surface comes from
the summation over $\vec{k}$ present in
$\Pi_{\nu}^{0}(\vec{q},\omega)$ and the assumption that all
scattering directions are equally probable. This is directly
connected with neglecting interactions between electrons and holes
within a bubble, and in particular the vertex correction within the bubble
coming from the Coulomb interaction. Given the long range nature of the
interaction, it is natural to expect that such corrections would enhance
the small $q$ scattering. We can thus expect that the averaged mass that
enters our expression for the dispersion relation is closer to the
mass along the direction of $q$ rather than the averaged mass.

Although a full calculation is difficult and clearly beyond the
scope of the present paper we can have an idea of the importance
of such effects by comparing to limiting situations, for the case
of $q$ along the trigonal axis. We take on one hand the averaging
of the eRPA\cite{Bennacer_anisotrop92} and on the other hand the
unidimensional approximation where we simply retain the mass along
$\vec{q}$. For an isotropic case the expressions
(\ref{eq:collecmode}) considerably simplify and allow to define
two important parameters for the equation namely the relative
weight between the two terms $R =
(\kappa_{T-F}^{s}/\kappa_{T-F}^{r})^2 =(m_s k_{Fs})/(m_r k_{Fr})$
(where $\kappa_{T-F}^{i}$ was defined in the context of
Eq.\ref{eq:potential_Kukk}) and $r_v = V_{Fs}/V_{Fr}$ which is the
ratio of the two velocities. The results are given in
Table.~\ref{tab:plasm_trig}.
\begin{table}
\begin{center}
\begin{tabular}{|c|c|c|c|c|c|}
  \hline
       & R & $r_{v}$ & $c_{\rm ac}$ & $\tau^{-1}$ & $k^*$ \\
  \hline
  eRPA & 11 & 0.2  & 1.8 & 0.4  & 0.25 \\
  UD   & 17 & 0.14 & 2.1 & 0.45 & 0.3 \\
  \hline
\end{tabular}
\end{center}
\caption{\label{tab:plasm_trig} The parameters R and $r_{v}$ (see
text for definition) of acoustic plasmon along the trigonal axis
in the limit $q\rightarrow 0$. eRPA is an average taking partially
into account the ellipsoidal shape of the Fermi surface, while UD
is a purely unidimensional approximation. The good agreement
between these two extreme cases suggests that the corresponding
results are probably representative of the experimental values in
Bismuth}
\end{table}
We see that two extreme approximations give rather similar values.
The correct values for Bismuth should therefore be between these
two extremes. This suggests that $k^{*}$ is far from being
negligible. Taking such values for $k^{*}$ gives an energy of
$T^{*} \approx 6K$ for the temperature below which the acoustic
plasmon becomes strongly coherent.

In the perpendicular plane the situation is significantly more
complex. In fact we have to work with a multi-plasma problem given
the three electron pockets so we can only give qualitative
arguments. Let us stay within the UD approximation, remembering
that it tends to slightly overestimate $k^{*}$. From
Fig.~\ref{fig:band-structure} we immediately see that by changing the
$\vec{q}$ direction we are facing the following situation:
\begin{itemize}
    \item there is one central hole pocket with an angle independent mass $m_{h}^{1,2}=0.067$.
    \item the mass of each electron pocket can vary drastically.
    from very light $m_{e}^{2}=0.0015$ up to the heaviest of all $m_{e}^{1}=0.198$
    \item the electron pockets are situated with an relative orientation of $\pi/3$ and
    thus for every $\vec{q}$ direction there are either very light and/or very heavy electrons present.
\end{itemize}
This open the possibility for several plasmons to emerge, but most
of them are strongly Landau damped. A rough estimation gives (for
$\vec{q}$ not far from the bisectrix between two pockets): for
light hole-heavy electron $R\approx 1.15 \Rightarrow k^{*} \approx
0.07$; for light electron-heavy hole $R\approx 1.2 \Rightarrow
k^{*} \approx 0.1$; for light electron-heavy electron $R\approx 13
\Rightarrow k^{*} \approx 0.22$.  For another intermediate
orientation ($\pi/6$ angle with the bisectrix) both electron
masses will be smaller. A more precise evaluation of $k^{*}$ would
be quite difficult and probably also susceptible to a significant
error. Let us only emphasize that for the Fermi energies
$E_{F}^{e}\approx 2 E_{F}^{h}$, thus a smaller value of $k^{*}$ in
the perpendicular plane does not immediately imply a smaller
$T^{*}$ in this plane.

Finally let us make two comments for plasmons far from the trigonal axis:
\begin{enumerate}
    \item If we had taken the eRPA approximation we would have found rather low values of
    $k^{*}$ along the bisectrix axis which would imply a notable trigonal/bisectrix anisotropy of the observed
    $T^{*}$. This is not the case found experimentally.
    \item Due to particular positioning of Fermi surface pockets
    it is practically always possible to find carriers with rather
    different masses. Thus we expect that a drop of $k^{*}$ for some directions is
    rather the exception than the rule. These rare cases should not
    affect significantly the $\vec{q}$ momentum integrals such as
    the one in (\ref{eq:res_result}).
\end{enumerate}

Note that in the above estimation of the $T^{*}$ values we have
neglected local field corrections. These always tend to reduce
$T^{*}$ \cite{Viginale_beyondRPA88}. Precisely: the larger $k^{*}$
the larger influence of local field corrections. The "Deybe"
temperatures found by resistivity fits by Uher\cite{Uher_Bi-rho77}
are thus lower but this possible discrepancy was already discussed
in the Sec.\ref{sec:new_rho}.

\section{Anisotropy of the Baber resistivity} \label{ap:ratioapp}

In order to compute the anisotropy of the resistivity due to the Baber scattering
we start from the formula obtained in Ref.~\onlinecite{Giamarchi_Shastry92} (Eq.~2.14
in this paper) and apply it to our problem with slow and rapid carriers:
\begin{multline}\label{eq:TG_Sh}
    \rho_{x}= \rho_{0}^{x} \beta \sum_{k,p,q}\nu_{pkq}^2
    f(\xi_{s}(k))(1-f(\xi_{s}(k)+\Delta_q))\\
    f(\xi_{r}(p))(1-f(\xi_{r}(p)-\Delta_q))
\end{multline}
where $\Delta_q$ is the energy exchanged in the scattering process
(determined by the exchanged momenta $q$) and $\rho_{0}^{x}=\pi
J^2/(\Omega D_{x})$, $D_{x}$ being the density of carriers in a
given direction, $J$ the strength of carrier's interaction. In our
model, instead of the $J$ of (\ref{eq:TG_Sh}), included in
$\rho_{0}$, the averaged scattering rate $W$ must be substituted.
Following the derivation of Ref.~\onlinecite{Giamarchi_Shastry92}
we take the carriers velocity along the resistivity axis as the
variational parameter $\Phi_{k^{i}}=V_{x}$. According to textbook
procedure (p.283 in Ref.\onlinecite{ziman_phonon_book})
$\nu_{pkq}$ is then equal to the total change of $\Phi_{k^{i}}$
during scattering event which in our case means $\nu_{pkq}=\Delta
V_{x}$. When one derives the relation $\Delta V_{x}(q)$, one needs
to take the masses along $x$-axis. The momentum along a given
direction must be conserved, which implies that $|\Delta
V_{x}(q)|\sim q_{x}|m_{ex}^{-1}-m_{hx}^{-1}|$ where $q_{x}$ is the
momentum exchanged along resistivity axis.

The summation in (\ref{eq:TG_Sh}) is taken over vectors of
incoming and outgoing carriers. The detailed, all-T treatment of
(\ref{eq:TG_Sh}) would be quite complicated, but we are
interested only in lowest temperatures, where the energy exchanged
during scattering goes to zero $\Delta\rightarrow 0$ (we always
stay on the Fermi surface) and the density of both electrons and
holes is constant (the chemical potential in semimetal is rather
stable). We also know from band structure that there is no nesting
$k_{e}\neq k_{h}$ and these Fermi wavevectors are also constant.

Note that the derivation in Ref.~\onlinecite{Giamarchi_Shastry92}
was done for a two dimensional system. Working in three dimensions
requires one extra angle $\vartheta$ to complete the spherical
coordinates. It is an angle between incoming momenta of light and
heavy carriers between $\angle(\vec{k},\vec{p})$ which was
obviously equal to zero for 2D case. As long as we are interested
in the resistivity along the high symmetry directions (which is
the one usually measured) and in the limit of small $\Delta$ we
may assume that the functional relations
$\Delta(\theta_{\vec{k}\vec{q}},\theta_{\vec{p}\vec{q}})$ and
$|q|(\theta_{\vec{k}\vec{q}},\theta_{\vec{p}\vec{q}})$,
originating from energy conservation, do not depend on the new
angle $\vartheta$. One integrates out this extra variable, but
this does not change the low energy $T$-dependence (because the
condition $\Delta\rightarrow 0$ allows to give a unique relation
between $\theta_{\vec{k}\vec{q}},\theta_{\vec{p}\vec{q}}$).

With the above remarks we can evaluate the
momentum sums, transposed into integral over the angles
$\theta_{\vec{k}\vec{q}},\theta_{\vec{p}\vec{q}}$, $\vartheta$.
The masses $m_{s}(\theta_{\vec{k}\vec{q}})$ and
$m_{r}(\theta_{\vec{p}\vec{q}})$ entering to formulas accounting
for energy conservation (during scattering event)
$\Delta(\theta_{\vec{k}\vec{q}},\theta_{\vec{p}\vec{q}})$ and
$|q|(\theta_{\vec{k}\vec{q}},\theta_{\vec{p}\vec{q}})$ are
averaged over all angles on the Fermi surface when integration
over all possible orientations in taken. This brings us to an
analog of Eq.~3.4 in Ref.~\onlinecite{Giamarchi_Shastry92} with the anisotropy factor
extracted:
\begin{multline}\label{eq:TGSh_simpl}
    \rho_{x}=\frac{(\Delta V_{x}/q_{x})^2}{D_x}\int d\theta_{\vec{k}\vec{q}}
    d\theta_{\vec{p}\vec{q}} d\vartheta q^{2}\cos^2(\vartheta)\\
    (\cos(\theta_{\vec{k}\vec{q}})-\cos(\theta_{\vec{p}\vec{q}}))^2\frac{(\Delta_{\vec{q}}\beta)^2}{4\sinh^2(\Delta_{\vec{q}}\beta/2)}
\end{multline}
where an integral is angle averaged which also means averaged over
all momenta on the Fermi surface. The integral is rather
complicated in 3D. We can safely assume that the anisotropy of the
resistivity is caused only by the two factors in front: a change
of carriers velocity $\Delta V_{x}$ (along the resistivity
direction) during the scattering event which we already discussed
above and the denominator $D_x$. The denominator $D_x$ is constant
when $\mu(T)=cste$ and it introduces the probability of such a
scattering event, which is proportional to carrier density
determined along the given direction by $D_x \sim
(m_{ex}^{-1}+m_{hx}^{-1})^2$. The resistivity can be expressed as
$\rho_x \sim |\Delta V_{x}|/D_x$. This implies that, in the low
temperatures (when $\rho\sim T^2$), the main contribution to the
anisotropy of resistivity is accounted by the factor $|\Delta
V_{x}|/D_x$ which after straightforward simplification gives:
\begin{equation}\label{eq:Baber_angle-dep}
    \rho_{i}(m_{ei},m_{hi})=\left[\frac{|m_{ei}-m_{hi}|}{(m_{ei}+m_{hi})}\right]^2\cdot I
\end{equation}
where $I$ is angle averaged integral. In the case of two pockets
Fermi surface this immediately leads to:
\begin{equation}\label{eq:ratioapp}
    \frac{\rho_{x}}{\rho_{z}}=\left[\frac{|m_{ex}-m_{hx}|/(m_{ex}+m_{hx})}{|m_{ez}-m_{hz}|/(m_{ez}+m_{hz})}\right]^2
\end{equation}
$m_{e}$ and $m_{h}$ are second rank tensor with values (along main
axis) given in Tab.~\ref{tab:masses}. A reminder is necessary: the
masses which are substituted into (\ref{eq:ratio}) are not the ones
given in Tab.~\ref{tab:masses}. Indeed the masses in (\ref{eq:ratio}) are
the ones along external directions while the Tab.~\ref{tab:masses}
gives the ellipsoid parameters. This point is particularly
important for electrons along the trigonal axis: there is very
small $m_{3e}$ but (due to out of bisectrix plane tilt $\vartheta$
of the electron ellipsoid) $m_{ez}$ contains an admixture of
relatively large $m_{1e}$:
\begin{equation}\label{eq:mass_comb}
    m_{ez}^{-1}=m_{3e}^{-1}\cos(\vartheta)^2+m_{1e}^{-1}\sin(\vartheta)^2
\end{equation}
There is one extra complication, not present only if the selected
direction is the trigonal one. It comes from the fact that in
Bismuth we have more than two families of carriers. In the
particular case of the bisectrix axis (along which resistivity is
frequently measured) we have to deal with three different types of
carriers (holes, light and heavy electrons) with certain
probabilities for each type of scattering event. This leads to:
\begin{equation}\label{eq:HTbis-res}
    \rho_{x}=\frac{3}{11}\rho_{1eh}(m_{e1},m_{h1})+\frac{6}{11}\rho_{2eh}(m_{e2},m_{h1})+\frac{2}{11}\rho_{ee}(m_{e1},m_{e2})
\end{equation}
which one can compare with $\rho_{z}$ measured along trigonal axis
precisely what we did in Sec.~\ref{ssec:standard_Baber}.


\begin{thebibliography}{10}

\bibitem{SdH_classic30}
L.~Schubnikov and W.J. de~Haas.
\newblock A new phenomenon in the change of resistance in a magnetic field of
  single crystals of bismuth.
\newblock {\em Nature}, 126:500, 1930.

\bibitem{dHvA_classic30}
W.J. de~Haas and P.M. van Alphen.
\newblock A new phenomenon in the change of magnetization in a magnetic field
  of single crystals of bismuth.
\newblock {\em Proc. Netherlands Roy. Acad. Sci}, 33:1106, 1930.

\bibitem{Tediosi_pressure}
N.~P. Armitage, Riccardo Tediosi, F.~L\'evy, E.~Giannini,
L.~Forro, and
  D.~van~der Marel.
\newblock Infrared conductivity of elemental bismuth under pressure: Evidence
  for an avoided lifshitz-type semimetal-semiconductor transition.
\newblock {\em Phys. Rev. Lett.}, 104(23):237401, Jun 2010.

\bibitem{Tediosi_plasmaron07}
Riccardo Tediosi, N.~P. Armitage, E.~Giannini, and D.~van~der
Marel.
\newblock Charge carrier interaction with a purely electronic collective mode:
  Plasmarons and the infrared response of elemental bismuth.
\newblock {\em Phys. Rev. Lett.}, 99(1):016406, Jul 2007.

\bibitem{Uher_Nernst_e-h78}
C.~Uher and W.~P. Pratt~Jr.
\newblock Thermopower measurements on bismuth from 9K down to 40 mK.
\newblock {\em Journal of Physics F: Metal Physics}, 8(9):1979, 1978.

\bibitem{Behnia_Nernst07}
Kamran Behnia, Marie-Aude M\'easson, and Yakov Kopelevich.
\newblock Nernst effect in semimetals: The effective mass and the figure of
  merit.
\newblock {\em Phys. Rev. Lett.}, 98(7):076603, Feb 2007.

\bibitem{Behnia_BiSb}
Aritra Banerjee, Beno\^\i{}t Fauqu\'e, Koichi Izawa, Atsushi
Miyake, Ilya
  Sheikin, Jacques Flouquet, Bertrand Lenoir, and Kamran Behnia.
\newblock Transport anomalies across the quantum limit in semimetallic
  $Bi_{0.96}Sb_{0.04}$.
\newblock {\em Phys. Rev. B}, 78(16):161103, Oct 2008.

\bibitem{Hartmann_experim-full69}
Robert Hartman.
\newblock Temperature dependence of the low-field galvanomagnetic coefficients
  of bismuth.
\newblock {\em Phys. Rev.}, 181(3):1070--1086, May 1969.

\bibitem{Kukkonen_experim77}
C~A Kukkonen and K~F Sohn.
\newblock The low-temperature electrical resistivity of bismuth.
\newblock {\em Journal of Physics F: Metal Physics}, 7(7):L193, 1977.

\bibitem{Kraak_highT82}
W.~Kraak, R.~Herrmann, and H.~Haupt.
\newblock Investigation of the charge carrier scattering in Bi under high
  hydrostatic pressure.
\newblock {\em phys. stat. sol. (b)}, 109(2):785--792, Jul 1982.

\bibitem{Kukkonen_Baber76}
C~A Kukkonen and P~F Maldague.
\newblock The electrical resistivity of bismuth:electron-hole scattering.
\newblock {\em Journal of Physics F: Metal Physics}, 6(11):L301, 1976.

\bibitem{Uher_Bi-rho77}
C.~Uher and W.~P. Pratt.
\newblock High-precision, ultralow-temperature resistivity measurements on
  bismuth.
\newblock {\em Phys. Rev. Lett.}, 39(8):491--494, Aug 1977.

\bibitem{Kukkonen_el-phon78}
Carl~A. Kukkonen.
\newblock $T^{2}$ electrical resistivity due to electron-phonon scattering on a
  small cylindrical fermi surface: Application to bismuth.
\newblock {\em Phys. Rev. B}, 18(4):1849--1853, Aug 1978.

\bibitem{Maude_graphite09}
J.~M. Schneider, M.~Orlita, M.~Potemski, and D.~K. Maude.
\newblock Consistent interpretation of the low-temperature magnetotransport in
  graphite using the Slonczewski-Weiss-McClure 3d band-structure calculations.
\newblock {\em Phys. Rev. Lett.}, 102(16):166403, Apr 2009.

\bibitem{Behnia_graphite09}
Zengwei Zhu, Huan Yang, Benoit Fauque, Yakov Kopelevich, and
Kamran Behnia.
\newblock Nernst effect and dimensionality in the quantum limit.
\newblock {\em Nat Phys}, 6:26, 2009.

\bibitem{Aebi_EI07}
H.~Cercellier, C.~Monney, F.~Clerc, C.~Battaglia, L.~Despont,
M.~G. Garnier,
  H.~Beck, P.~Aebi, L.~Patthey, H.~Berger, and L.~Forr\'o.
\newblock Evidence for an excitonic insulator phase in $1T-TiSe_{2}$.
\newblock {\em Phys. Rev. Lett.}, 99(14):146403, Oct 2007.

\bibitem{Tutis_SC09}
A.~F. Kusmartseva, B.~Sipos, H.~Berger, L.~Forr\'o, and
  E.~Tuti\ifmmode~\check{s}\else \v{s}\fi{}.
\newblock Pressure induced superconductivity in pristine $1T-TiSe_{2}$.
\newblock {\em Phys. Rev. Lett.}, 103(23):236401, Nov 2009.

\bibitem{Baber_classics37}
W.~G. Baber.
\newblock {The Contribution to the Electrical Resistance of Metals from
  Collisions between Electrons}.
\newblock {\em Proceedings of the Royal Society of London. Series A -
  Mathematical and Physical Sciences}, 158(894):383--396, 1937.

\bibitem{Allen_Bi-band95}
Yi~Liu and Roland~E. Allen.
\newblock Electronic structure of the semimetals Bi and Sb.
\newblock {\em Phys. Rev. B}, 52(3):1566--1577, Jul 1995.

\bibitem{Ted_thesis}
R.~Tediosi.
\newblock {\em Pressure tuning of low-energy collective excitations in metals}.
\newblock PhD thesis, Universite de Geneve, 2008.

\bibitem{Pines_ac-plasm-classic62}
David Pines and J.~Robert Schrieffer.
\newblock Approach to equilibrium of electrons, plasmons, and phonons in
  quantum and classical plasmas.
\newblock {\em Phys. Rev.}, 125(3):804--812, Feb 1962.

\bibitem{Ruvalds_ac-plasm-review81}
J.~Ruvalds.
\newblock Are there acoustic plasmons?
\newblock {\em Advances in Physics}, 30(5):677, 1981.

\bibitem{Cottey_ac-plasm-basic85}
A.~A. Cottey.
\newblock Acoustic plasmons in a two-component fermi gas.
\newblock {\em Journal of Physics F: Metal Physics}, 15(8):L203, 1985.

\bibitem{Bennacer_ac-plasm-Bi89}
B.~Bennacer, A.~A. Cottey, and J.~Senkiw.
\newblock Calculated acoustic plasmon spectra in GaSb, SnTe and Bi.
\newblock {\em Journal of Physics: Condensed Matter}, 1(45):8877, 1989.

\bibitem{Das_Sarma_ac-plasm99}
S.~Das~Sarma and E.~H. Hwang.
\newblock Collective charge-density excitations in two-component
  one-dimensional quantum plasmas: Phase-fluctuation-mode dispersion and
  spectral weight in semiconductor quantum-wire nanostructures.
\newblock {\em Phys. Rev. B}, 59(16):10730--10743, Apr 1999.

\bibitem{Fenton_rhoBi67}
E.~W. Fenton, J.~P. Jan, \AA{}. Karlsson, and R.~Singer.
\newblock Ideal resistivity of bismuth-antimony alloys and the
  electron-electron interaction.
\newblock {\em Phys. Rev.}, 184(3):663--667, Aug 1969.

\bibitem{Giamarchi_Shastry92}
T.~Giamarchi and B.~Sriram Shastry.
\newblock Baber scattering and resistivity of a two-dimensional two-band model.
\newblock {\em Phys. Rev. B}, 46(9):5528--5535, Sep 1992.

\bibitem{mahan_book}
G.~D. Mahan.
\newblock {\em Many-Particle Physics}.
\newblock Physics of Solids and Liquids. Kluwer Academic/Plenum Publishers, New
  York, third edition, 2000.

\bibitem{Whitfield_polaron}
George Whitfield and P.~M. Platzman.
\newblock Simultaneous strong and weak coupling in the piezoelectric polaron.
\newblock {\em Phys. Rev. B}, 6(10):3987--3992, Nov 1972.

\bibitem{Noto_magnetoplasm74}
Hiromi Noto.
\newblock Collective modes of an electron-hole system in a magnetic field.
\newblock {\em Journal of the Physical Society of Japan}, 36(4):1137--1147,
  1974.

\bibitem{Dirkgroup}
D~van~der Marel, Riccardo Tediosi, and F.~L\'evy.

\bibitem{PhysRevE.62.5648}
H.~Reinholz, R.~Redmer, G.~R\"opke, and A.~Wierling.
\newblock Long-wavelength limit of the dynamical local-field factor and
  dynamical conductivity of a two-component plasma.
\newblock {\em Phys. Rev. E}, 62(4):5648--5666, Oct 2000.

\bibitem{PhysRevB.53.9797}
B.~Adolph, V.~I. Gavrilenko, K.~Tenelsen, F.~Bechstedt, and
R.~Del~Sole.
\newblock Nonlocality and many-body effects in the optical properties of
  semiconductors.
\newblock {\em Phys. Rev. B}, 53(15):9797--9808, Apr 1996.

\bibitem{Abrikosov_BiRG73}
A.A Abrikosov.
\newblock The electrical resistivity of bismuth:electron-hole scattering.
\newblock {\em Journal of Low Temperature Physics}, 10(1):1, 1973.

\bibitem{Norin_Bi(p)DFT77}
Brage Norin.
\newblock Temperature and pressure dependence of the band structure in bismuth.
\newblock {\em Physica Scripta}, 15(5-6):341, 1977.

\bibitem{Balla_Bi(p)65}
D.~Balla and B.~Brandt.
\newblock {\em Sov. Phys.-JETP}, 20:1111, 1965.

\bibitem{Mendez_Bi(p)81}
E.~E. Mendez, A.~Misu, and M.~S. Dresselhaus.
\newblock Pressure-dependent magnetoreflection studies of Bi and
  $Bi_{1-x}Sb_{x}$ alloys.
\newblock {\em Phys. Rev. B}, 24(2):639--648, Jul 1981.

\bibitem{Uher_graphite84}
D.~T. Morelli and C.~Uher.
\newblock $T^{2}$ dependence of the in-plane resistivity of graphite at very
  low temperatures.
\newblock {\em Phys. Rev. B}, 30(2):1080--1082, Jul 1984.

\bibitem{Wilson_TiSe-diel78}
J.~A. Wilson, A.~S. Barker, F.~J.~Di Salvo, and J.~A.
Ditzenberger.
\newblock Infrared properties of the semimetal Ti$Se_{2}$.
\newblock {\em Phys. Rev. B}, 18(6):2866--2875, Sep 1978.

\bibitem{Levy_rhoTiSe79}
F.~Levy.
\newblock Electrical resistivity and hall effect in $TiSe_2$ containing vanadium
  impurities.
\newblock {\em Journal of Physics C: Solid State Physics}, 12(18):3725, 1979.

\bibitem{Bennacer_anisotrop92}
B~Bennacer and A~A Cottey.
\newblock Calculated acoustic plasmon spectra in SnTe: effect on anisotropy.
\newblock {\em Semiconductor Science and Technology}, 7(6):822, 1992.

\bibitem{Viginale_beyondRPA88}
G.~Vignale.
\newblock Acoustic plasmons in a two-dimensional, two-component electron
  liquid.
\newblock {\em Phys. Rev. B}, 38(1):811--814, Jul 1988.

\bibitem{ziman_phonon_book}
J.~M. Ziman.
\newblock {\em Electrons and Phonons}.
\newblock Clarendon, Oxford, 1962.

\end{thebibliography}

\end{document}